%% file: main.tex
\documentclass[lettersize,journal]{IEEEtran}
\usepackage{amsmath,amsfonts}
\usepackage{algorithmic}
\usepackage{algorithm}
\usepackage{array}
\usepackage[caption=false,font=normalsize,labelfont=sf,textfont=sf]{subfig}
\usepackage{textcomp}
\usepackage{stfloats}
\usepackage{url}
\usepackage{verbatim}
\usepackage{graphicx}
\usepackage{cite}
\usepackage{xcolor}
\usepackage{multirow}
\usepackage{orcidlink}  % For Orcid
\newcommand{\merge}[1]{\textcolor{black}{#1}}
\newcommand{\nxt}[1]{\textcolor{black}{#1}}
\newcommand{\final}[1]{\textcolor{black}{#1}}

\newcommand{\quan}[1]{\textcolor{black}{#1}}

\newcommand{\tuan}[1]{\textcolor{black}{#1}}

\newcommand{\TODO}[1]{\textcolor{black}{#1}}
\newcommand{\rebuttal}[1]{\textcolor{black}{#1}}
\newcommand{\debug}[1]{\textcolor{black}{#1}}  %% Consistency Check
\begin{document}

\title{PEAT: Pseudo-Error Assessment for GPU Kernel Validation in DNN Training}

\author{Xuan Truong Nguyen\orcidlink{0000-0002-7527-6971}, Hong Quan Tran, Tran Thang Le, Tuan Duc Chu, and Thanh Tuan Dao\orcidlink{0000-0002-9897-2769}
\thanks{This work was supported by Moreh and Van-Lang Institute of Semiconductor Technology (VIST). (\textit{Corresponding author: Thanh Tuan Dao}.)

Xuan Truong Nguyen is with the Department of Next Generation Semiconductor Convergence and Open Sharing System (COSS), Seoul National University, Seoul 08826, South Korea. He is also with Van-Lang Institute of Semiconductor Technology (VIST), Hanoi, Vietnam.
%, Hanoi, Vietnam (E-mail: truongnx@snu.ac.kr).

Hong Quan Tran, Tuan Duc Chu, Tran Thang Le, and Thanh Tuan Dao are with the Efficient Computation Research Group, Vietnam National University. Thanh Tuan Dao is also with Moreh Vietnam. %, Hanoi, Vietnam (E-mail: \{quan.tran, duc.chu, tuan.dao\}@moreh.com.vn).
} 
}

\maketitle
% % As a general rule, do not put math, special symbols or citations
% % in the abstract or keywords.
\input{sections/0_Abstract}
\input{sections/1_Introduction}
\input{sections/2_Background}
\input{sections/3_Motivation}

\input{sections/4_Method}
\input{sections/5_Characterization}

\input{sections/6_Related_works}
\input{sections/7_Conclusion}



\bibliographystyle{IEEEtran}
\bibliography{references}
\end{document}

%% file: sections/0_Abstract.tex
\begin{abstract} \label{sec:abstract}
Deep neural networks (DNNs) are widely adopted in various fields, driving an emerging trend in developing software stacks associated with DNN training systems. For example, many codes have been ported across different frameworks or developed to leverage the computing power of GPUs or domain-specific accelerators. However, validating a kernel implementation in DNN training is time-consuming and generally requires massive storage. Specifically, this poses a fundamental question: how to characterize the behavior of a new implementation when it is integrated into a DNN training flow. Unfortunately, this problem is not well investigated in the literature, to the best of our knowledge.

To address this shortcoming, we present PEAT - a lightweight inspection framework for \underline{P}seudo-\underline{E}rror \underline{A}ssessment associated with GPU kernel validation in DNN \underline{T}raining. 
Firstly, inspired by conventional fault injection (FI), PEAT's Profiler invokes an operation-wise kernel in a training flow to collect a DNN model's states (e.g., checkpoints and activations). More importantly, the Profiler introduces two simple yet effective techniques, playback FI and frequency-based runtime FI, leveraging persistent kernel calling during the training process. Secondly, PEAT's Analyzer characterizes profiled errors, revealing some signatures from the error distribution of a kernel compared to the golden one. Lastly, PEAT's Detector provides some guidelines as a sufficient condition, which enables associating several well-known error models with signature patterns.
%PEAT invokes an operation-wise kernel in a training flow to collect the states of a DNN model (e.g., checkpoints and activations). Next, a custom kernel is injected into the training context (named playback fault injection), effectively visualizing error patterns such as convergence and repetition. %We demonstrate the applicability of our approach with NVIDIA V100 and AMD MI250 GPUs on various DNN models for both pretraining and fine-tuning scenarios.}
\tuan{We demonstrate the applicability of our approach by presenting the results and analysis using GPUs from the two most popular vendors, NVIDIA V100 and AMD MI250, on various AI models, from vision tasks to language models, for both pretraining and finetuning scenarios.}
\end{abstract}


%% file: sections/1_Introduction.tex
\section{Introduction} \label{sec:Introduction}

%% Paragraph 1: DNN training and kernel development
Deep neural network (DNN) training workloads have become increasingly prevalent in data centers. As DNNs trend toward larger and more complicated models, the DNN training process is becoming more and more \tuan{resource-intensive~\cite{le-scao-etal-2022-language, golden2024flashattentionstable}}. For example, training the LLaMA2’s 70B parameter model required 1,720,320 GPU hours \cite{touvron2023llama2openfoundation}. 
\tuan{
On the other hand, the fast-paced development of the AI infrastructure industry (in both hardware and software) has posed a challenge to DNN framework developers to evolve their code.
As a result, \quan{developers} must invest significant effort in developing new code or porting and optimizing existing code to support emerging hardware architectures~\cite{furiosa_tcp, SambaNova, Intel_Gaudi}.
%constantly optimize their codes to improve their performance and port their existing codes to support new hardware architectures~\cite{furiosa_tcp, SambaNova, Intel_Gaudi}. 
}
\debug{For example, many approaches utilize quantization that exploits data types with a low bit width such as half-precision (FP16) and BFloat16 (BF16) \cite{kalamkar2019_bfloat16}, FP8 \cite{Wang_2018_fp8, micikevicius_2022_fp8formatsdeeplearning}, dynamic floating-point \cite{das2018mixed}, block floating-point \cite{Drumond_2018_hybrid_block_FP, Qian_hpca22_fast}, flex-point \cite{Koster_2017_flexpoint}, or fixed-point \cite{das2018mixed}. 
From a hardware perspective, modern GPUs support various data types such as FP32, FP16, INT8, and INT4~\cite{nvidia_v100, nvidia_a100}. Meanwhile, another approach to speed up DNN training is to develop new GPU code, such as operation fusions or FlashAttention \cite{Tri_Dao_FlashAttention}.}
\tuan{These activities, combined with the complexity of both hardware and software architectures, make GPU code validation particularly challenging.
}

%%% NOTE: Simplify Introduction 
%To speed up the training process on GPUs, a common approach is to increase GPU utilization and throughput and minimize data transfers to off-chip DRAM.
%The most direct way to reduce off-chip or external memory accesses (EMA) is quantization by using data types with fewer bits per weight or activation. For example, half-precision (FP16) and BFloat16 (BF16) \cite{kalamkar2019_bfloat16} directly enable a bandwidth saving of 50\% compared to FP32. Relying on domain-specific accelerators (DSA), low-bit data types can be extended to FP8 \cite{Wang_2018_fp8, micikevicius_2022_fp8formatsdeeplearning}, dynamic floating-point \cite{das2018mixed}, block floating-point \cite{Drumond_2018_hybrid_block_FP, Qian_hpca22_fast}, flex-point \cite{Koster_2017_flexpoint}, or fixed-point \cite{das2018mixed}. From a hardware perspective, modern GPUs support various data types such as FP32, FP16, INT8, and INT4~\cite{nvidia_v100, nvidia_a100}
%\TODO{cite Nvidia, AMD}
%, enabling accelerating DNN training via mixed-precision training. Meanwhile, one approach to speed up DNN training is to develop a new GPU code such as operation fusions or Flash Attention \cite{Tri_Dao_FlashAttention}. For example, Flash Attention uses tiling to reduce the number of memory reads/writes between a high bandwidth memory (HBM) on a GPU's card and GPU's on-chip SRAM.

\begin{figure}[t!]
\centering
\includegraphics[width=1.0\columnwidth]{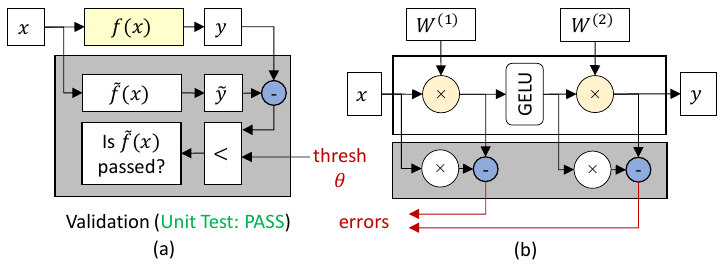}
%\vspace{-10pt}
\caption{Example of Kernel Validation at a Unit Test (a) and a Potential Accuracy Drop at a System Test (b).}
\label{fig01:intro_accuracy_wall}
\end{figure}

%% Kernel validation
\debug{There is a substantial body of research that focuses on GPU code verification~\cite{Betts_2012_GPUVerify, Betts_2015_GPU_Verify, Kamath_ISCA_Data_race, Kamath_2021_iGUARD, wu2020simulee, sorensen2021specifying}. 
However, previous work primarily uses formal methods to identify software bugs, typically concurrency issues. A key underlying assumption of this approach is that bugs can be identified by analyzing the outcomes of device code execution.
Although this approach is effective in detecting \tuan{generic} data races and synchronization bugs, it is not sufficient to reveal silent errors in DNN training~\cite{Ma_asplos24_drDNA}.}
\tuan{Also, the emphasis on identifying individual error sources makes it impractical to account for the errors that accumulate across different parts in the code.}

\begin{figure}
\includegraphics[width=1.0\columnwidth]{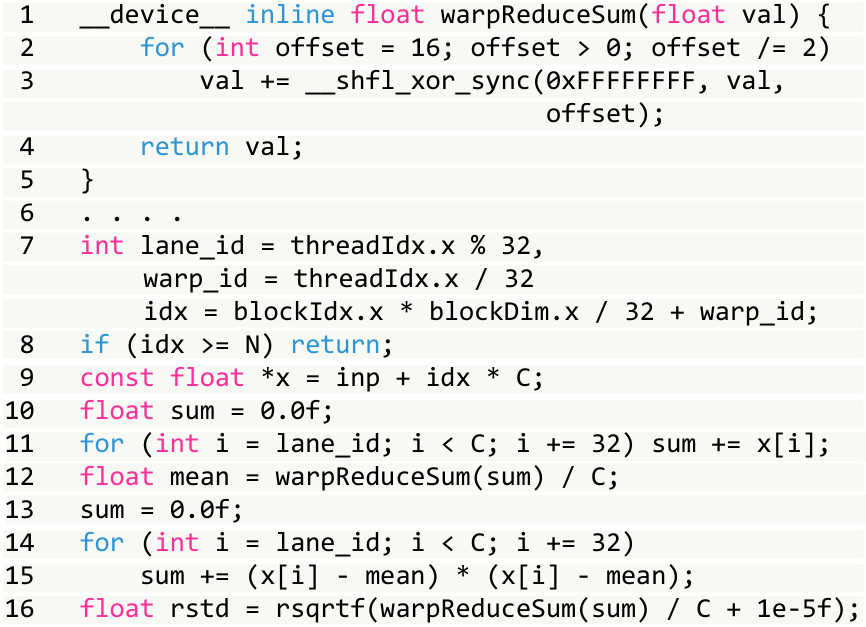}
\caption{\TODO{A layer normalization implementation calculating a summation sequentially (e.g., lines 10$-$11 or lines 13$-$15) which introduces an "error",  compared to a default Pytorch one.}}
\label{fig:def-layernorm}
\end{figure}

\begin{figure}[t!]
\includegraphics[width=1.0\columnwidth]{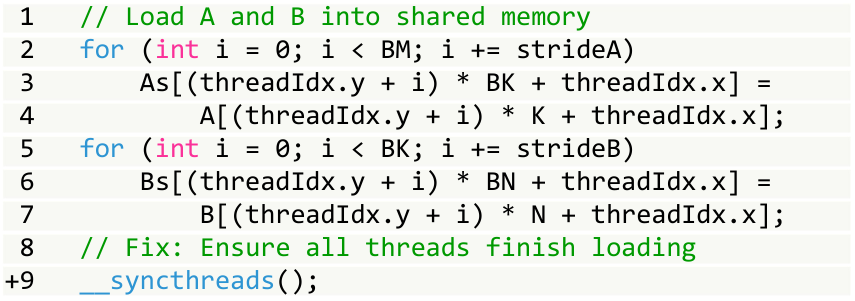}
\caption{Data Race Example. An error occurs when \texttt{syncthreads} is missed.}
\label{fig:def-race}
\end{figure}

\debug{To validate a new device code (e.g., GPU kernel), a common approach is to compare its output against golden implementations of the well-known DNN frameworks such as PyTorch~\cite{paszke2019_pytorch} and TensorFlow~\cite{tensorflow2015-whitepaper}, as illustrated in Fig.~\ref{fig01:intro_accuracy_wall}(a). 
However, the development may introduce a tiny error in code modification, for example, precision discrepancies in a GEMM kernel with FP16~\cite{IEEE_754} compared to its FP32 counterpart. 
\rebuttal{These \tuan{tiny} errors refer to a small change in an output result of a new code compared with that of a reference version (often called \textit{golden}). Tiny errors are a class of errors with a small yet noticeable magnitude (for example, \tuan{from $~10^{-4}$ to $10^{-2}$} in the context of DNN training) during a unit test. A common source of errors comes from floating-point computations, as illustrated in Fig.~\ref{fig:def-layernorm}. While a new kernel is semantically correct, a different execution order of floating-point operations introduces a small error compared with the golden kernel. Another typical source of errors comes from changing the code semantics (software bugs), for example, a data race bug as illustrated in Fig.~\ref{fig:def-race}.}}

\debug{Analyzing such a tiny numerical deviation is increasingly essential for Large-Language Models (LLMs), where training time may take months across hundreds or thousands of GPUs. 
By the nature of very long training time, \emph{training instability} has become increasingly problematic. 
As reported with Google's PaLM model, training instability often manifests itself in loss spikes that occur up to 20 times throughout training~\cite{google_2020_palm}. 
These loss spikes often lead to expensive operations as they usually cause interruptions in the training process, requiring training to stop and restart. 
Small parameter perturbations (e.g., parameter updates) can be applied (so-called \emph{amplification effect}~\cite{liu-etal-2020-understanding}), which may result in significant disturbances in the model output or even training crashes. Recently, a numerical deviation was also analyzed~\cite{golden2024flashattentionstable} to answer the question if a recent effective flash attention \cite{Tri_Dao_FlashAttention} is stable during training.}

\debug{These tiny errors present a challenge for developers in determining the semantic correctness of the modified code. As shown in Fig.~\ref{fig01:intro_accuracy_wall}, kernel validation typically relies on a user-defined threshold $\theta$. 
However, determining a proper $\theta$ is a nontrivial task in kernel integration for DNN training. 
For example, a GEMM kernel can be used in down and up projecting operations in a transformer layer, as illustrated in Fig.~\ref{fig01:intro_accuracy_wall}(b), resulting in different error ranges. 
Consequently, a threshold $\theta$ may fail to reflect diverse error ranges in a DNN training context.
Recent works have investigated errors in GPU codes. 
For example, they have explored the floating-point arithmetic implemented in the NVIDIA tensor cores~\cite{nvidia_v100, nvidia_turing}, and the Ampere microarchitecture~\cite{nvidia_a100}. 
They specifically design numerical experiments to understand the difference between the results produced by code for tensor coders~\cite{fasi_2021_tensor_cores}. 
Since realistic inputs to a GPU code tend to have distributions that differ significantly from those of theoretically randomized inputs, this may lead to theoretical bounds underestimating or overestimating the precision errors.}

\debug{Another challenge to characterize the error source in kernel validation is monitoring cost. 
For example, the output must be monitored and observed for a "sufficiently long" period to notice the error accumulation when an error, which is injected into a new code, is generally tiny.
Unfortunately, the massive amount of execution of the GPU code makes the cost of monitoring any part of the DNN training very expensive in terms of both computing resources and storage. 
This problem becomes much more severe for DNNs with hundreds of billions of \tuan{parameters}.}

%\debug{To address the aforementioned challenges, this work presents PEAT, a lightweight pseudo-error assessment for GPU kernel validation during DNN training with the following features.}
To address the above challenges, this work presents PEAT, a lightweight pseudo-error assessment for GPU kernel validation during DNN training with the following key features.
\begin{itemize}
    \item \debug{\textbf{Context-Aware Profiling.} PEAT introduces two simple yet effective concepts, \emph{playback FI} and \emph{frequency-based runtime FI}. These FI schemes specifically focus on characterizing an error (referred to as a pseudo-error) caused by the injection of a device code during DNN training. Unlike other fault injection approaches, PEAT introduces a new error profile scheme by gathering errors across multiple epochs \tuan{with a reasonable cost}. }
    \item \debug{\textbf{Error Range Characterization.} Based on the obtained error profiles, PEAT reveals a "convergence" pattern that has been unseen before. Consequently, PEAT adopts common metrics, such as mean and variance, on an error profile, which enables the observation of various error types across different scenarios, including operation-wise, model-wise, and injection position-wise.}
    \item \debug{\textbf{Error Classification.} Leveraging diverse range characterization, PEAT suggests a simple error classifier that categorizes an error in kernel validation by its range. For example, it is observed that injecting a GEMM kernel during a forward pass may cause much larger errors (e.g., a few order) than that with the backward pass.}
\end{itemize}

\merge{The paper is organized as follows. Section~\ref{sec:Background} briefly reviews some background, including DNN operations and errors on DNNs. The motivation and the proposed PEAT are described in Sections~\ref{sec:Motivation} and~\ref{sec:PEAT}, respectively. The evaluation results are presented in Section~\ref{sec:Characterization}. Section~\ref{sec:conclusion} draws a conclusion.}

%% file: sections/2_Background.tex
\section{Background} \label{sec:Background}

%\TODO{Explain DNN operations => Emphasize DNN Blocks with identical structures}
\subsection{\nxt{DNNs, Model Scaling, and Framework Development}} \label{subsec:DNN Model Scaling}
%\debug{\textbf{Operations in DNNs.} CNN models~\cite{He2015_resnet, tan2020_efficientnet, Andrew_MBv1} consist of primitives such as convolution (CONV), batch normalization (BN)~\cite{IoffeS15_batchnorm}, pooling, and fully-connected (FC) layers, as illustrated in Fig.~\ref{fig:DNN Models}(a). Transformer-based LLMs such as \cite{google_2020t5, google_2020_palm} typically consist of multiple encoder or decoder blocks. Each block includes multi-head attention (MHA), feed-forward networks (FFN), layer normalization (LN)~\cite{ba2016layernormalization}, and residual connections~\cite{He2015_resnet}.} 

\begin{figure}[t!]
\centering
\includegraphics[width=0.95\columnwidth]{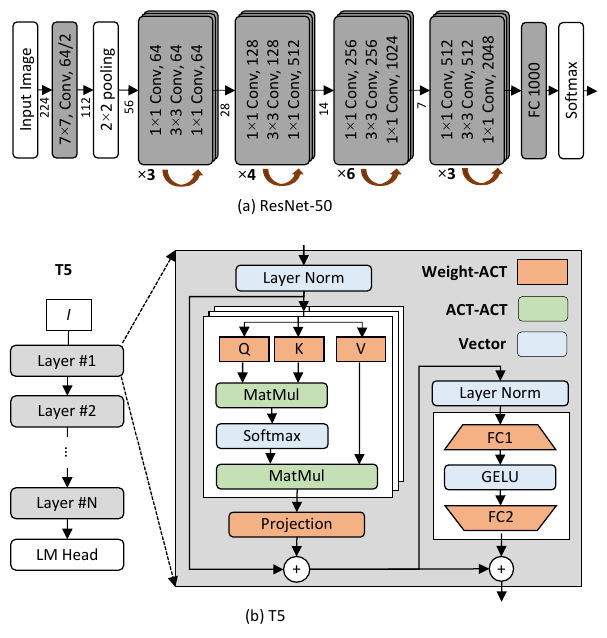}
\vspace{-5pt}
\caption{\nxt{Repetitive Blocks in (a) ResNet-50 and (b) T5.}}
\label{fig:DNN Models}
\end{figure}

\nxt{\textbf{DNNs.} Convolutional neural network models~\cite{He2015_resnet, tan2020_efficientnet, Andrew_MBv1} consist of primitives such as convolution (Conv), batch normalization (BN)\cite{IoffeS15_batchnorm}, pooling, and fully-connected (FC) layers, as illustrated in Fig.~\ref{fig:DNN Models}(a). 
A Conv layer calculates the output activations by multiplying the input activations by weight tensors, and it is typically followed by a BN layer.
FC layers may be included at the end of a CNN model for several specific tasks such as image classification. 
%For certain tasks, such as image classification, an FC layer is typically invoked at the end of a CNN model.
}%\nxt{\textbf{Large Language Models (LLMs).} 
Meanwhile, transformer-based LLMs such as \cite{google_2020t5, google_2020_palm} typically consist of multiple encoder or decoder blocks. Each block includes multi-head attention (MHA), feed-forward networks (FFN), layer normalization~\cite{ba2016layernormalization}, and residual connections~\cite{He2015_resnet}. 
A given input token is first processed by layer normalization, and the normalized result is multiplied with three weight matrices to generate query (Q), key (K), and value (V). In MHA, Q, K, and V are divided into multiple heads to compute attention scores, probabilities, and outputs within each head. The outputs of multiple heads are concatenated and projected by an FC layer. Meanwhile, FFNs include two FC layers and produce expanded intermediate results during the process.
%}

\nxt{\textbf{Model Scaling.} Many DNNs, including ResNet \cite{He2015_resnet}, MobileNet \cite{Andrew_MBv1}, and EfficientNet \cite{tan2020_efficientnet} \tuan{use} multiple blocks with the same structure
at different image scales
, as illustrated in Fig.~\ref{fig:DNN Models}(a). 
For example, six CNN blocks are used at the feature size of 14. 
At this image scale, ResNet-101 and ResNet-152 adopt 23 and 36 blocks, respectively. 
Similarly, T5~\cite{google_2020t5} also uses multiple encoder blocks, for example, 6 and 24 for T5-small and T5-3B models. \tuan{
Repetitive structure is commonly used to scale DNN architecture~\cite{He2015_resnet, tan2020_efficientnet, Andrew_MBv1, google_2020t5, google_2020_palm}. 
}
This suggests a strong message that porting or developing a kernel, for example, Conv in ResNet \cite{He2015_resnet} or FC in T5 \cite{google_2020t5}, may largely affect DNN models.}

\nxt{\textbf{DNN Framework Development.}} 
\merge{Modern DNN training systems (Pytorch~\cite{paszke2019_pytorch}, Pytorch-rocm, TensorFlow~\cite{tensorflow2015-whitepaper}) and DNN inference systems (TVM~\cite{Tianqi_tvm}, vLLM~\cite{kwon2023efficient}, MIGraphX~\cite{MIGraphX}) typically incorporate multiple GPUs and a system software stack, including a compiler and runtime~\cite{Tianqi_tvm, pldi21_Jung, taco2025_lee, micro22_wei, ispass20_Ananda}. Users commonly work on domain-specific applications via popular DL frameworks such as PyTorch~\cite{paszke2019_pytorch} or TensorFlow~\cite{tensorflow2015-whitepaper}. 
Meanwhile, new hardware architectures~\cite{furiosa_tcp, SambaNova, Intel_Gaudi, asplos24_seo} also introduce new operations, which typically raise a strong demand for system code development and optimization.
%With the fast-paced development of AI workloads, new DNN architectures~\cite{furiosa_tcp, SambaNova, Intel_Gaudi} often introduce new operations. The development of various target hardware architectures may demand an emerging trend of code development and optimization.
}

\subsection{\nxt{Kernel Validation and Errors in DNNs}} \label{subsec:Error Models}

\nxt{\textbf{Kernel Code with Tiny Errors.}} 
\merge{A tiny error is highly related to precision errors due to floating-point computations, which is vastly studied in literature~\cite{Castaldo_siam08, blanchard_2019, fasi_2021_tensor_cores, Arar_siam23}. However, these approaches barely consider a practical DNN training context. For example, the proposed Playback FI reveals that the error mean of a convolution code tends to change in a small range during DNN training, which is underexplored in the conventional theoretical analysis~\cite{Castaldo_siam08, blanchard_2019, fasi_2021_tensor_cores, Arar_siam23}.
}

\nxt{\textbf{Error Characterization, Detection, and Mitigation.}} 
\merge{Soft errors have received a vast of attention recently in both DNN training systems~\cite{He_isca2023_HWFI, Guerrero_sc_23} and inference systems~\cite{Ma_asplos24_drDNA, Kamath_ISCA_Data_race, Bolchini_sc23, Singh_sc23}. A system fault during DNN training is widely known to be hard to detect. For example, according to ~\cite{He_isca2023_HWFI}, in 82.3\%$-$90.3\% of all cases across the workloads, the injected faults did not significantly affect the final training/test accuracy for the same training time as the fault-free runs. 
In fact, the majority of them (65.5\%$-$86.3\% of all cases) yielded slightly higher training/test accuracy compared to the fault-free cases. More importantly, it is costly to characterize a fault during DNN training, for example, $>$2.9 M ($>$490 K node hours) FI experiments~\cite{he_micro20_fidelity}.
%PEAT, on the other hand, offers a low-complexity solution to investigate a new kernel under a DNN training context.
}

%\textbf{Device Code Development.}
%There is a substantial body of research that focuses on GPU code verification~\cite{Betts_2012_GPUVerify, Betts_2015_GPU_Verify, Kamath_ISCA_Data_race, Kamath_2021_iGUARD, wu2020simulee, sorensen2021specifying}. 
%However, previous work primarily uses formal methods to identify software bugs, typically concurrency issues. A key underlying assumption of this approach is that bugs can be identified by analyzing the outcomes of device code execution.
%While this approach is effective to detect data races and synchronization bugs, it falls short of revealing silent errors in DNN training~\cite{Ma_asplos24_drDNA}. 

%To validate a new device code (e.g., GPU kernel), developers typically compare its output against golden implementations from the well-known DNN frameworks such as PyTorch~\cite{paszke2019_pytorch} and TensorFlow~\cite{tensorflow2015-whitepaper}. 
%However, the development may introduce a tiny error in code modification, for example, precision discrepancies in a GEMM kernel with FP16~\cite{IEEE_754} compared to its FP32 counterpart.
%These tiny errors introduce a challenge for developers in determining the semantic correctness of the modified code.

\nxt{\textbf{Error Models.} Several common and practical error models consist of random bit flip (RBF) \cite{he_micro20_fidelity, Reagen_DAC18_ARES_BitFlip}, random value (RV) \cite{he_micro20_fidelity}, random value - stealthy, and random Not-a-Number (NaN). RBF is a widely acknowledged error model emulating transient hardware errors \cite{he_micro20_fidelity, Reagen_DAC18_ARES_BitFlip}. This error model is designed to simulate the flipping of network parameters such as weights given a bit error rate. This RBF may cause a random NaN. For example, consider a bitflip case with a 32-bit single precision (FP32) or a 16-bit half-precision (FP16) numbers as regulated by IEEE-754 standard \cite{IEEE_754}. The number in the range between 1.0 and 2.0 will have an eight-bit exponent as "0111\_1111" in the FP32 format. In this case, flipping the first bit from "0" to "1" results in an NaN number. RV is another widely used error model to describe the parameter value variations inside a neural network \cite{he_micro20_fidelity}. Specifically, the value of a parameter may change into a random one based on a given probability. In realistic scenarios, the range of values, i.e., weights or activations, in a layer may vary in a small range of possible values. For example, due to a casting from FP32 to F16, some small values may become zero (so-called a random zero (RZ) error).} 
%\subsection{\truong{Error Models}}
%\truong{Several common and practical error models consist of random bit flip (RBF) \cite{he_micro20_fidelity, Reagen_DAC18_ARES_BitFlip}, random value \cite{he_micro20_fidelity}, random value - stealthy, and random Not-a-Number (NaN). RBF is a widely acknowledged error model emulating transient hardware errors \cite{he_micro20_fidelity, Reagen_DAC18_ARES_BitFlip}. This error model is designed to simulate the flipping of network parameters such as weights given a bit error rate. This RBF may cause a random NaN. For example, consider a bitflip case with a 32-bit single precision (FP32) or a 16-bit half-precision (FP16) numbers as regulated by IEEE-754 standard \cite{IEEE_754}. The number in the range between 1.0 and 2.0 will have an eight-bit exponent as "0111\_1111" in the FP32 format. In this case, flipping the first bit from "0" to "1" results in an NaN number. RV is another widely used error model to describe the parameter value variations inside a neural network \cite{he_micro20_fidelity}. Specifically, the value of a parameter may change into a random one based on a given probability. In realistic scenarios, the range of values, i.e., weights or activations, in a layer may vary in a small range of possible values. For example, due to a casting from FP32 to F16, some small values may become zero (so-called a random zero (RZ) error).}

\textbf{Resilience Analysis and Fault Injection.} 
Resilience analysis is essential for understanding the behavior of hardware errors in the development of DNNs and DL accelerators, especially in safety-critical applications.
A common method to observe the error behavior is to inject different types of errors into a golden code and record the outcomes.
Several popular error injection tools have been proposed, including FIdelity~\cite{he_micro20_fidelity}, PyTorchFI~\cite{Mahmoud_2020_PytorchFI}, ARES~\cite{Reagen_DAC18_ARES_BitFlip}, BinFI~\cite{Chen_SC19_BinFI}, and GoldenEye~\cite{GoldeneyeMahmoudTambeDSN2022}. 
Inspired by flip-flop (FF) faults in a hardware system, He {et al.}~\cite{He_isca2023_HWFI} presents an error case where a bit flip error may be injected for a couple of cycles. 
However, these frameworks have several limitations. 
For example, they have limited support for diverse network architectures like Transformer-based models. 
While they can configure an error rate in a specific tensor, their settings are not well-suitable for modeling pseudo-persistent errors. 
For example, they do not account for error models that result from a faulty GPU kernel being called multiple times during the training phase.

%% file: sections/3_Motivation.tex
\section{Motivation} \label{sec:Motivation}
%\subsection{Difficulty in Optimized Kernel Development}
%\TODO{Limitations of naive test cases in validating a buggy kernel}

%% NXT - DEL

%\begin{figure}[t!]
%\centering
%\includegraphics[width=1.0\columnwidth]{fig/Fig02_FI_Kernel_Injections.pdf}
%\vspace{-10pt}
%\caption{\nxt{Example of Precision and Programming Errors in a Kernel (a) and Injected Kernels in DNN Blocks (b).}}
%\label{fig:FI_Kernels}
%\end{figure}

This section presents several new challenges and opportunities in characterizing a kernel's behavior during DNN training. 
%These are mainly related to the unique characteristics of kernel validation - a kernel is typically invoked multiple times (so-called \emph{persistent kernel call (PKC)} during a training flow.

\subsection{Challenges} \label{sec:Motvation_challenges}
\textbf{Error attribution.} 
To verify the semantic correctness of a given kernel, developers often rely on the training loss or the validation \tuan{loss} as the most important metrics.
However, discrepancies in these metrics may result from hardware or software, each with its own variety of error types.
This makes pinpointing the root causes of incorrect results particularly challenging. 
Identifying these errors requires a deep understanding of not only the hardware architecture and software stack but also the AI application's behavior.

\textbf{Tiny errors.} 
A key task in developing a new kernel is performing unit tests. 
Developers generally compare the output of the new kernel to a golden implementation or ground-truth. 
If the difference falls within a "sufficiently small" range, \tuan{the kernel is considered to be correct.}
%the test is considered to be successful.
However, determining a pass condition can be challenging when the output contains \textit{tiny errors}.
Several factors contribute to this difficulty.
First, DNNs can tolerate a wide range of precision levels without significantly affecting their accuracy~\cite{Wang_2018_fp8,pmlr-v162-huang22h,das2018mixed,evans2021acgc,kalamkar2019_bfloat16,micikevicius_2022_fp8formatsdeeplearning}.
Second, most computations in DNNs are performed using floating-point format (e.g., FP32 or FP16), where precision depends on the underlying implementation, hardware architecture, and the training data.
\tuan{
Third, the numerous layers in DNNs lead \quan{to} substantial variations in value ranges across layers~\cite{dong2019hawq,kim2023squeezellm}, necessitating different pass criteria for each layer. 
}

\begin{figure}[t!]
\centering
\includegraphics[width=0.85\columnwidth]{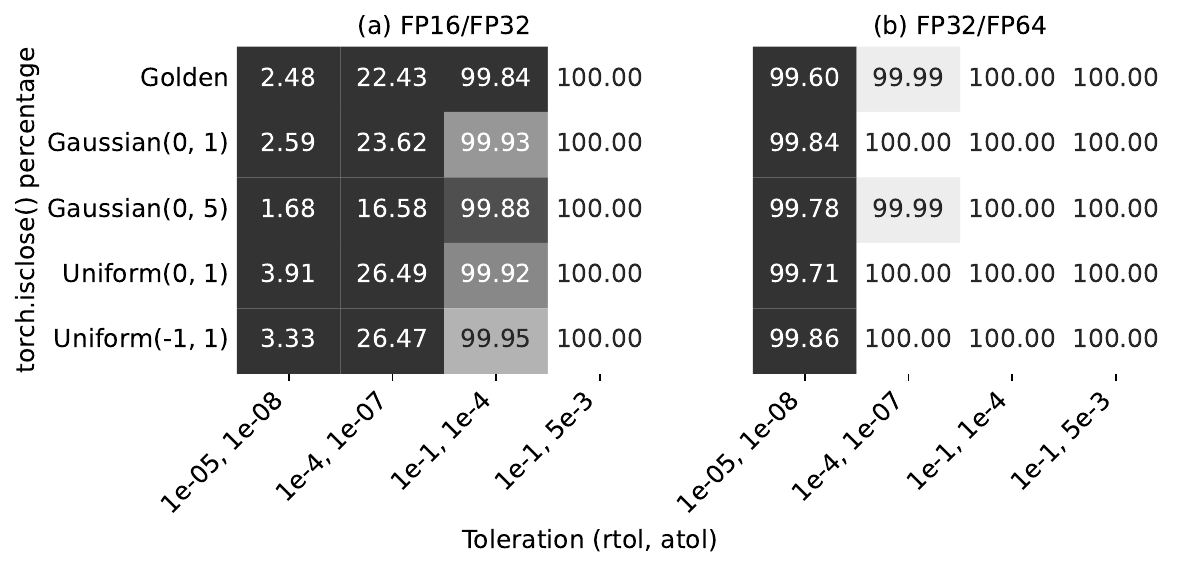}
\vspace{-5pt}
\caption{Percentages of output elements of a convolution operation for which \texttt{torch.isclose()} returns \texttt{true}.}
\label{fig05:Motivating_isclose}
\end{figure}

Figure~\ref{fig05:Motivating_isclose} demonstrates a simple example where the tiny errors can create issues. 
In this experiment, we execute a convolution operation using two pairs of precision formats (FP16-FP32 and FP32-FP64) on the same input and calculate the number of output elements that are deemed "sufficiently close". 
The \textit{closeness} is determined using the function \texttt{torch.isclose} with different tolerance thresholds \texttt{(rtol, atol)}.
We show the percentage of output elements for which \texttt{torch.isclose} returns \texttt{true} for FP16 and FP32 in Fig.~\ref{fig05:Motivating_isclose}(a) and for FP32 and FP64 in Fig.~\ref{fig05:Motivating_isclose}(b).

\begin{figure}[t!]
\centering
\includegraphics[width=0.9\columnwidth]{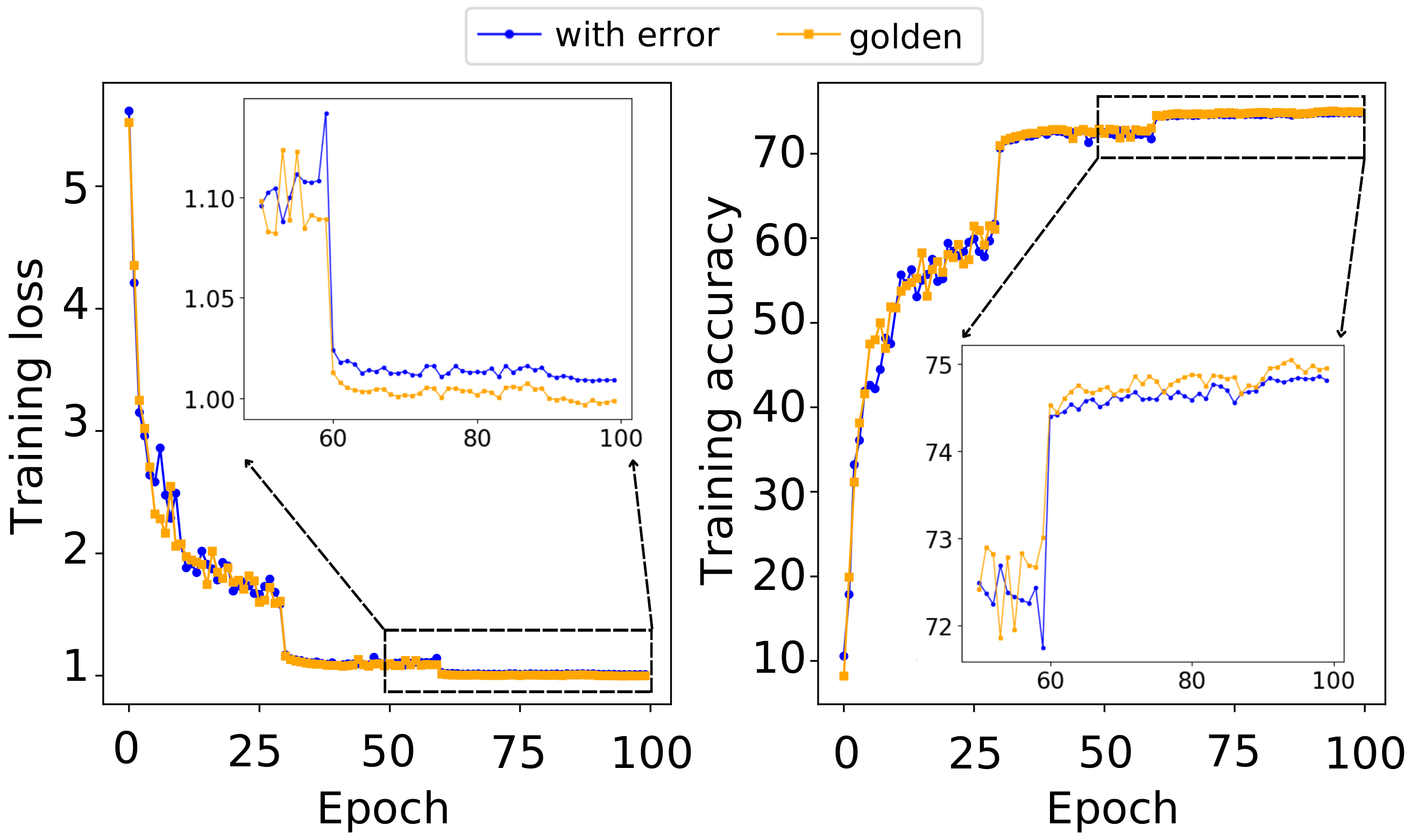}
\vspace{-10pt}
\caption{\tuan{An accuracy drop case with Resnet-50 on ImageNet. The \textit{error} version illustrates the model training with a kernel that contains a concurrency bug.}}
\label{fig:Accuracy_drop_ImageNet}
\end{figure}

The most significant observation from the results is that, despite performing the same algorithm (2D convolution), the number of output elements that fall within the specified range differs greatly when we vary the input distribution and tolerance values.
Moreover, this number also differs when comparing different pairs of precision formats (FP16 vs. FP32 and FP32 vs. FP64).
The interplay of input distribution, tolerance, and precision clearly complicates the kernel validation process within DNN frameworks.
\rebuttal{Additionally, a faulty kernel may cause an accuracy drop, as illustrated in Fig.~\ref{fig:Accuracy_drop_ImageNet}. The \quan{experiment} is conducted with Resnet-50 on the ImageNet dataset. It is observed that a faulty kernel with a concurrency bug slightly \quan{increases the training loss and reduces} training accuracy. \tuan{The faulty kernel has an error range of $10^{-4}$ and may \quan{pass} most of kernel validation routine in general.}}

\textbf{Large characterization space.} 
To characterize the behavior of errors in large DNNs, a common approach is to inject faults into the execution of a kernel of interest.
However, due to both the DNN architecture and the training algorithm, a kernel is typically executed multiple times during the training process.
The location (in a DNN architecture) and timing (including time steps, training iterations, and epochs) at which errors are injected can significantly affect the outcomes and may lead to different insights.
The number of combinations of locations and times for error injection is extremely large, making it generally intractable to explore all possibilities.
Tackling this challenge requires a strategy to carefully select where and when to inject errors.

%Another challenge for kernel validation is related to the huge number of scenarios for a system test. Figure~\ref{fig03:FI_Kernels}(b) illustrates two feed-forward layers (\circled{4}, (\circled{6}))and a GELU layer (\circled{5}) in a common Transformed-based model. Considering a case in which each layer can be implemented by one among four implementations shown in Figure~\ref{fig03:FI_Kernels}(a), we will have a total of $81(=3^4)$ options. In practice, a network may include a few hundred layers, resulting in a huge number of cases for kernel validation. Additionally, due to model scaling, DNN models typically consist of many structure-identical blocks, as explained in Section~\ref{subsec:DNN Model Scaling}, leading to an injection of a kernel at different locations. These pose a strong demand to design a pragmatic approach for validating a new implementation at a system level. %\TODO{May explain time and storage complexity.}

%\begin{figure}[t!]
%\centering
%\includegraphics[width=0.90\columnwidth]{fig/Fig04_Motivating_Example_T5.pdf}
%\vspace{-10pt}
%\caption{\nxt{Example of accuracy difference between Golden and Fault-Injection cases in T5 on ConLL2003 dataset.}}
%\label{fig04:Motivating_Example_T5}
%\end{figure}

%\begin{figure}[t!]
%\centering
%\includegraphics[width=0.90\columnwidth]{fig/Fig04_Motivating_Example_Effnet.pdf}

%\caption{\nxt{Example of accuracy difference between Golden and Fault-Injection cases in EfficientNet on CIFAR10 dataset.}}
%\label{fig04:Motivating_Example_EfficientNet}
%\end{figure}

\subsection{Opportunities} \label{sec:Motvation_oppportunities}
The challenges discussed in the previous section suggests that characterizing the kernel behavior requires a broader context than solely analyzing the kernel itself.
This presents the opportunities to provide a richer context based on several factors in DNN training.
Unlike conventional fault injection (FI) models, FI in a DNN training flow offers control over when, where and how long a kernel is injected. 
We refer to the position where a kernel is injected as the \emph{Fault Injecting Point (FIP)}. 
This capability allows us to monitor the effect of errors, such as their location, time, and amplitude.
The position where the error is observed or monitored is termed as the \emph{Fault Observing Point (FOP)}.

%\textbf{Training Context.} 
The training context refers to the information that can be gathered during the DNN training phase.
To effectively analyze kernel behavior within this context, several factors should be considered. First, a specific kernel implementation is often called multiple times in a training iteration. This suggests that if the error occurs at one location, it is likely to occur at other locations as well. Second, The same kernel is called repeatedly for each iteration. If an error occurs, it may propagate to subsequent iterations and potentially be observed later in the training process. Third, DNN architectures often have a repeating block of similar structures. This structural similarity provides a valuable hint for identifying error patterns.
%\begin{enumerate}
%   \item A specific kernel implementation is often called multiple times in a training iteration. This suggests that if the error occurs at one location, it is likely to occur at other locations as well.
%    \item The same kernel is called repeatedly for each iteration. If an error occurs, it may propagate to subsequent iterations and potentially be observed later in the training process.
%    \item DNN architectures often have a repeating block of similar structures. This structural similarity provides a valuable hint for identifying error patterns.
%\end{enumerate}

By properly setting the FIP and FOP, we can evaluate the immediate impact of error injection (when FIP equals FOP) or analyze the propagated impact after multiple iterations or epochs (when FIP is smaller than FOP). 
This approach differs from typical unit testing processes, which typically validate a kernel on limited synthetic or practical data.

%% file: sections/4_Method.tex
\section{PEAT} \label{sec:PEAT}
%%\TODO{1. Draw the overview PEAT: Profiler, Analyzer, and Detector }
%This section presents PEAT, which consists of a Profiler (Fig.~\ref{fig:PEAT_profiler}), Analyzer (Fig.~\ref{fig:PEAT_analyzer})
%%, and Detector
%to inspect a GPU kernel in a DNN training framework. 
%Inspired by conventional fault injection (FI), the Profiler invokes an operation-wise kernel in a training flow to collect a DNN model's states (e.g., checkpoints and activations). 
%More importantly, the Profiler introduces two simple yet effective techniques, playback FI and frequency-based runtime FI, leveraging persistent kernel calling (PKC) during the training process. 
%The Analyzer characterizes profiled errors, revealing signatures and patterns from the error distribution of a kernel compared to the golden one. 
%Lastly, the Detector provides some guidelines to detect the errors of the PKC in the DNN training.
%%These result patterns can be used as guidelines to detect the errors in the PKC.

\debug{This section presents the methodology of PEAT. The building blocks in PEAT, including a Profiler, an Analyzer, and a Detector, are also introduced.}

\subsection{Methodology}
\label{sec:method}

%\textbf{Assumptions.} 
%In this work, we assume that a golden kernel is an implementation in a single precision FP32~\cite{IEEE_754}. 
%This represents a practical scenario when a kernel is executed on a device, for example, a GPU's CUDA cores. 
%It is widely known that FP32 computations also introduce a rounding error during floating-point operations~\cite{IEEE_754}. 
%Therefore, in our setting, we also use a double-precision (FP64) implementation as a way of injecting an error. 
%Notably, an FP32-FP64 error represents a rounding error occurring during the execution of the golden FP32 kernel.

\debug{
PEAT focuses on errors injected into fundamental operations in DNN models, such as convolution (Conv), batch normalization (BN), fully connected (FC), and layer normalization (LN) layers.  
Specifically, for Conv and FC layers, an error in a device code is emulated by corrupting weights, activations, or gradients. 
This scenario is common in kernel development, where a kernel associated with model parameters may introduce a tiny error due to bugs or optimizations. 
We assume that a golden kernel is an implementation in a single precision FP32~\cite{IEEE_754}.
This represents a practical scenario when a kernel is executed on a device, for example, a GPU's CUDA cores. 
It is widely known that FP32 computations also introduce a rounding error during floating-point operations~\cite{IEEE_754}. 
Therefore, in our setting, we also use a double-precision (FP64) implementation as a way of injecting an error. 
Notably, an FP32-FP64 error represents a rounding error occurring during the execution of the golden FP32 kernel.
}

\debug{
PEAT adopts several common pseudo-error types in kernel code development, such as FP16, random value (RV), and random zero (RZ). 
An FP16 error occurs by casting an operand tensor from single precision (FP32) to half-precision (FP16). 
This is a typical case where a model can be stored in FP16 instead of FP32, reducing the storage and external memory accessed by 50\%. 
On the other hand, RV and RZ errors represent the possibility of data corruption in a tensor.
The number of corrupted positions is controlled by a hyperparameter $\texttt{num\_pos}$. 
These errors aim to emulate CUDA programming bugs when a few threads may introduce an error, such as data races, leading to unexpected input values. 
When a tensor is corrupted, its corresponding output introduces an error (commonly referred to as an output tensor error) compared to that of the golden kernel. 
For example, Err(FP16, FP32) refers to an error in the output tensor when an FP32 input tensor is corrupted by casting it to FP16.
}

\begin{figure}[t!]
\centering
\includegraphics[width=0.95\columnwidth]{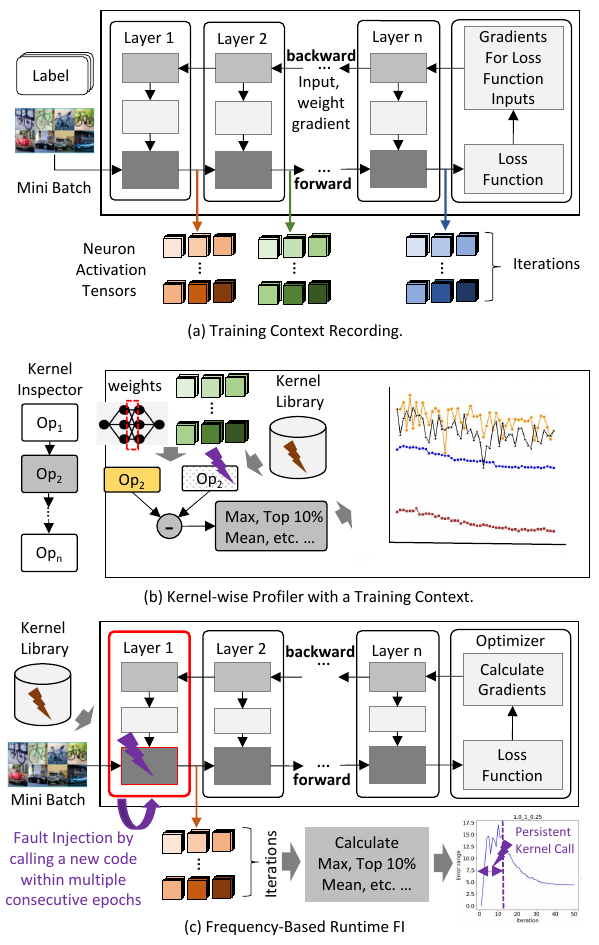}
\vspace{-10pt}
\caption{\nxt{PEAT's Profiler.}}
\label{fig:PEAT_profiler}
\end{figure}

\subsection{Profiler}
\subsubsection{Playback Fault Injection (PFI)}
\debug{
PEAT introduces a new playback fault injection (PFI) scheme to profile errors by calling a new kernel code. 
This includes two phases: recording and fault injection. 
During the recording stage, this records weights, activations, and gradients associated with the golden model during training, as illustrated in Figure~\ref{fig:PEAT_profiler}. 
The pre-recorded data serves as a training context. 
Next, in the fault injection stage, PFI invokes a "faulty" kernel on the pre-recorded data at a specific kernel or location. 
For example, given a set of 400 pairs (weight, activation), the PFI casts them into half-precision and then calls a kernel code to compute the output. 
An error profile is generated by comparing the computed results and the gold ones. Since the output is a generally four-dimension tensor, including height, width, channel, and batch, we empirically utilize a mean of top-10\% as a metric, which enables us to capture the trend of maximum errors with less noise.  
}

\debug{
PKI signifies two distinguished features compared to the conventional FI methods such as~\cite{Ma_asplos24_drDNA, He_isca2023_HWFI}. 
Like a typical unit test, it allows setting a fault-observing point as a fault-injecting point at the kernel. 
More importantly, unlike in ~\cite{Ma_asplos24_drDNA}, one important feature in PFI is that an error profile includes errors across multiple consecutive epochs, capturing the changes in weights, activations, and gradients during training, as illustrated in Figure~\ref{fig:PEAT_profiler}(b). Meanwhile, it is worth noting that PFI is cost-effective for performing various kernel validations within the same training context. 
For example, given pre-recorded data of a Resnet-50 training profile, it enables us to obtain an error profiler when injecting a Conv or BN kernel at various locations. 
A new kernel or a faulty kernel emulated by various data corruptions, such as FP16 and FP64 casting or random zero and random value injection, can also be tested with the same pre-recorded data. It is suggested that PKI is much more cost-effective than the conventional FI methods such as~\cite{He_isca2023_HWFI}. 
}

\TODO{
The RFI algorithm is summarized in Alg.~\ref{alg:playbackfi}. %TODO: Explain the algorithm line by line.
The algorithm takes activations ($A_i$), weights ($W_i$), and output gradients ($\delta O_i$) as inputs (\textit{line 1}). Notably, the tensors are recorded across multiple epochs, for example, $N$, during training. In the playback stage, it then performs $N$ iterations, where in each iteration, the “golden” results for both forward and backward operations are computed: the output ($O_i$), the activation gradients ($\delta A_i$), and the weight gradients ($\delta W_i$) (\textit{line 4}).
For precision fault injection, the computation uses casted recorded FP32 weights and activations into a lower precision, for example, FP16, and the outputs and gradients are compared against the golden values. For a programming fault injection, one or a few points in one of the three inputs ($A_i$, $W_i$, or $\delta O_i$) are corrupted by forcing them to random values or zeros injected with a fault (\textit{line 8}). The algorithm then computes the resulting outputs, activation gradient, and weight gradient based on the faulty input, as explained \textit{lines 9$-$11}, respectively.
The final error is defined as the difference between the golden results and those obtained after fault injection (\textit{lines 13$-$14}).
}

\subsubsection{Frequency-Based Runtime Fault Injection (RFI)}
\debug{
PEAT introduces another profiling scheme, frequency-based runtime fault injection (RFI), that is specifically oriented to device code validation. 
Unlike the playback fault injection, RFI simulates an error or bug in a kernel code by iteratively corrupting data across multiple training epochs, as shown in Fig.~\ref{fig:PEAT_profiler}(c). 
RFI offers two distinguished features compared to~\cite{He_isca2023_HWFI}.
First, leveraging the context of kernel validation, RFI determines both a fault injecting point and a fault observing point during training.
Like the playback FI, it extracts an error profile besides training accuracy or loss.
Second, unlike hardware errors ~\cite{He_isca2023_HWFI}, it enables us to continuously observe an error at the output of an injected kernel across consecutive kernel calls. 
This runtime fault injection compensates the cost-effective playback FI, monitoring progressive degradation and therefore providing a deeper understanding of how sustained error effects influence training. Although it is less cost-effective than PFI, it can serve as an intensive test to inspect a kernel at a particular injection point in advance. 
}

\TODO{
The RFI algorithm is summarized in Alg.~\ref{alg:runtimefi_final}.
Runtime FI begins by selecting the fault injection type: for forward or backward operations, or precision, as shown in \textit(lines 3\-11). For Precision FI, activations and weights are downcasted to FP16 (line 10), similar to the Playback FI. 
For the other three cases, which are collectively called typical programming FI cases in which a fault is injected into activation ($A$), weight ($W$), or output gradient ($\delta O$) (\textit{lines 4, 6, and 8}). 
Training then proceeds with the injected tensors, and at each iteration, activations, weights, and gradients are recorded (\textit{line 12}). Each case produces a modified set of tensors (\textit{lines 16$-$17}), which are compared against the golden results to compute the final error (\textit{line 19}), similar to the Playback FI.
}

\begin{algorithm}[tb!]
   \footnotesize
   \caption{Playback Fault Injection}
   \label{alg:playbackfi}
\begin{algorithmic}[1]
   \STATE {\bfseries Input:} Recorded tensors $\{(A_i, W_i, \delta O_i)\}_{i=1}^N$
    \FOR{$i = 1$ \textbf{to} $N$}
        \STATE \texttt{// Golden results}
        \STATE $(O_i, \delta A_i, \delta W_i) \leftarrow \text{FwdBwd}(A_i, W_i, \delta O_i)$
        
        \STATE \texttt{// Precision FI}
        \STATE $O_{fp16}, O_{fp64} \leftarrow \text{Fwd}_{fp16/fp64}(A_i, W_i)$
        
        \STATE \texttt{// Programming FI}
        \STATE $A'_i, W'_i, \delta O'_i \leftarrow \text{Inject}(A_i, W_i, \delta O_i)$
        \STATE $(O_{fwd\_a}, O_{fwd\_w}) \leftarrow (\text{Fwd}(A'_i, W_i),\ \text{Fwd}(A_i, W'_i))$
        \STATE $(\delta A_{gout}, \delta A_{w}) \leftarrow (\text{InGrad}(W_i, \delta O'_i),\ \text{InGrad}(W'_i, \delta O_i))$
        \STATE $(\delta W_{gout}, \delta W_{a}) \leftarrow (\text{WgGrad}(A_i, \delta O'_i),\ \text{WgGrad}(A'_i, \delta O_i))$
    \ENDFOR
    \STATE $ E \leftarrow\text{Err}(\{O_i, \delta A_i, \delta W_i\}, \{O_{fp16}, O_{fp64}, O_{fwd\_a}, O_{fwd\_w},$
    \STATE \hspace{10.5em}$\delta A_{gout}, \delta A_{w}, \delta W_{gout}, \delta W_{a}\})$

\end{algorithmic}
\end{algorithm}

\begin{algorithm}[tb!]
    \footnotesize
   \caption{Runtime Fault Injection}
   \label{alg:runtimefi_final}
\begin{algorithmic}[1]
   \STATE {\bfseries Config:} fault type $\in$ \{fwd, input\_grad, weight\_grad, precision\}
  
   \FOR{each epoch}
       \IF{fault type is fwd}
           \STATE $A$, $W \leftarrow \text{Inject}(A), W$ \textbf{or} $A$, $\text{Inject}(W)$
       \ELSIF{input\_grad}
           \STATE $\nabla O$, $W \leftarrow \text{Inject}(\nabla O), W$ \textbf{or} $\nabla O, \text{Inject}(W)$
       \ELSIF{weight\_grad}
           \STATE $\nabla O$, $A \leftarrow \text{Inject}(\nabla O), A$ \textbf{or} $\nabla O, \text{Inject}(A)$
        \ELSIF{precision}
            \STATE $\quad A, \text{Model} \leftarrow \text{to(FP16)}$
       \ENDIF
       \STATE Continue training, and record $(A, W, \nabla O)$
   \ENDFOR
    \STATE $\quad A, \text{Model} \leftarrow \text{to(FP32)}$
   \FOR{$i = 1$ to number of recorded samples}
       \STATE $(O_i, \nabla A_i, \nabla W_i) \leftarrow \text{FwdBwd}(A_i, W_i, \nabla O_i)$
        \STATE $(O^{\text{FI}}_i, \nabla A^{\text{FI}}_i, \nabla W^{\text{FI}}_i) \leftarrow \text{FwdBwd}(A^{\text{FI}}_i, W^{\text{FI}}_i, \nabla O^{\text{FI}}_i)$
   \ENDFOR
   \STATE $E \leftarrow \text{Err}(\{O, \nabla A, \nabla W\}, \{O^{\text{FI}}, \nabla A^{\text{FI}}, \nabla W^{\text{FI}}\})$
\end{algorithmic}
\end{algorithm}

\subsubsection{FI Assessment Scores}
To analyze the impact of fault injection on output precision, we compute several metrics to assess the error characteristics between the golden and fault-injected outputs, given that the error is defined as their absolute difference. 
These metrics provide a comprehensive view of deviation patterns and potential sensitivity to faults.
\begin{itemize}
    \item \textit{Max Error:} Captures the highest absolute error in a recorded tensor at an iteration, offering insight into peak deviations that may indicate worst-case fault impacts.
    \item \textit{Mean Error:} Represents the average deviation in a recorded tensor at an iteration, summarizing the general impact of faults on the output accuracy.
    \item \textit{Mean of Top 10\% Errors:} Focuses on the mean of top$-$10\% errors in a recorded tensor at an iteration, emphasizing the most significant deviations.
    \item \textit{Variance of Errors:} Provides the variance of errors, highlighting how faults affect output precision.
\end{itemize}
    
\subsection{Analyzer}
\subsubsection{Model-wise Error Range}
Operator-wise errors are recorded at locations and epochs across the training process, also known as a profile. However, the number of errors becomes extremely large when analyzing errors in a model. For example, for a popular Resnet-50 model with 107 layers, recording a layer-wise error in 400 epochs results in more than 40,000 data points. 
%To assess and compare the error ranges among different layer and locations, we introduce the mean ($\mu$) and standard deviation ($\sigma$) of errors across multiple epochs. 
%To represent the model-wise error range derived from these operator-wise errors, we compute their mean ($\mu$) and standard deviation ($\sigma$).
%We consider the errors between FP32 and FP64 as \textit{error units} caused by floating-point computations. 
%As a result, at each epoch, we obtain a relative error by dividing an absolute error by the corresponding FP32-FP64 error.
%In PEAT, an operator-wise error record consists of the errors obtained at different epochs to consider a real training context strictly. 
As a result, a single operation or layer can be represented by multiple epoch-based errors for a given error model. This presents a challenge in characterizing models at the model-wise level, for example, among different operations or layers at a specific location within repetitive blocks. To address this obstacle, we propose to use a pair of (mean, std) obtained from the errors. The simple metric enables us to capture the error range variation of a code during training. 

The PEAT's analyzer considers three levels of characterization: 
(1) A single operation in an entire model, 
(2) different operations, 
and (3) a group of operations in repetitive blocks. 
At the first level, it enables the characterization of a specific operation type, for example, Conv and BN layers in CNN models or FC and LN layers in Transformer models. 
Next, by considering different operations together, the Analyzer can compare and reveal signature patterns according to error models and operator types. 
Finally, the PEAT's analyzer characterizes signature patterns associated with repetitive blocks, including residual blocks in ResNet models~\cite{He2015_resnet} and encoder/decoder blocks in Transformer models~\cite{touvron2023llama2openfoundation}.

\subsubsection{Rank-based Error-Data Correlation}
To provide more insights into error patterns, the PEAT Analyzer also examines the correlation between errors and some widely-known criteria, such as weight mean and variance or activation mean and variance. 
We select the \textit{mean of the top 10\% errors} as the main score for further analysis. 
We use \textbf{Spearman correlation} to quantify the relationships between the scores of an FI error and a particular metric. 
The Spearman method captures rank-based associations, offering a robust measure of monotonic relationships that might not be linear. 
This approach is suitable given the diverse nature of these metrics and allows us to evaluate how well variations in FI errors relate to underlying model parameters and behaviors. 
The metrics include the means and variances of the weight and activation tensors.
%The detailed metrics are outlined as follows.
%We select the \textit{Mean of the top 10\% errors} as the main metric for further analysis. To investigate possible relationships, we calculate its correlation with the following model-specific metrics.

%\begin{itemize}
%    \item \textit{Weight Metrics:} We use the mean and variance of the original weight values, which may %indicate the sensitivity of specific weights to faults.
%    \item \textit{Activation Metrics:} Mean and variance of activation values offer insights into how varying %activation levels might correlate with error levels.
%    %\item \textit{Delta Weight:} The change in weight values across iterations can show if it coincides with %fault propagation.
%    %\item \TODO{\textit{Mean Exponent of Activation/Weight:} This metric reflects the magnitude and %distribution of activation and weight values.}
%    %\item \TODO{\textit{Quantization Error Metrics:} Calculated as the absolute difference when casting %activations and weights to FP16 and back to FP32, reflecting precision shifts.}
%\end{itemize}

\subsection{Detector}
Based on the profiled and analyzed results, the Detector reveals signature patterns featuring the fundamental operations such as convolution (CONV), linear, and batch normalization (BN) layers in DNN models as well as FCs and layer normalization in Transformer models. 
Similar to other detectors~\cite{Ma_asplos24_drDNA, Guerrero_sc_23}, our detector also compares scores against predefined thresholds. 
One score comprises the maximum and minimum values of means and variances, denoted $[xmin, xmax]$ and $[rmin, rmax]$, respectively, during a training context for a given DNN model. 
Another score is based on the Spearman metric described in the Analyzer.

\begin{figure}[t!]
\centering
\includegraphics[width=0.95\columnwidth]{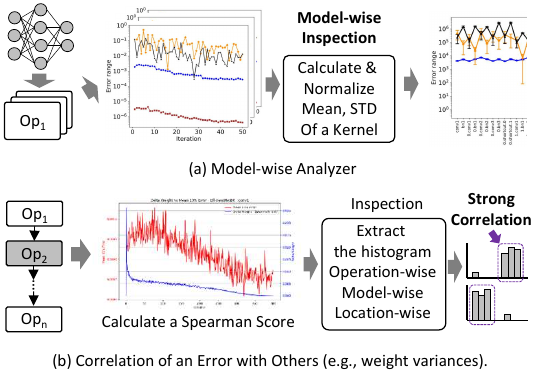}
\vspace{-10pt}
\caption{\nxt{PEAT's Analyzer.}}
\label{fig:PEAT_analyzer}
\end{figure}

%% file: sections/5_Characterization.tex
\section{Characterization} \label{sec:Characterization}

%% Divide the papers into three main parts: 
%% 1. Profiler -> Context-aware Playback FI + Runtime FI -> Similar to other FI campaigns but different. 
%% 2. Analyzer -> A tool to analyze errors => Scale-wise, Block-wise, Model-wise, Operation-wise
%% 3. Detector -> Pattern Recognition Guideline -> Signature patterns to characterize errors
%%      + Precision => Playback FI + Analyzer => Mean + variance
%%      + Random Zero => Playback FI + Normalization 
%%      + Random Value => Runtime FI with different ranges

%\begin{table}[h!]
\begin{table}[b!]
\centering
\footnotesize
\caption{\nxt{CNN and Transformer Workloads.}}
\label{tab:training_models}
 %\resizebox{0.8\columnwidth}{!}{
\begin{tabular}{lcl}
\hline
\textbf{Models}      & \textbf{Datasets}                       & \textbf{\# of layers}       \\ \hline
\multicolumn{1}{l}{EfficientNetB0}   & \multicolumn{1}{c}{\multirow{7}{*}{CIFAR-10}} & \multicolumn{1}{l}{131} \\  
\multicolumn{1}{l}{ResNet18}   & \multicolumn{1}{c}{}                  & \multicolumn{1}{l}{41} \\  
\multicolumn{1}{l}{ResNet34}   & \multicolumn{1}{c}{}                  & \multicolumn{1}{l}{73} \\  
\multicolumn{1}{l}{ResNet50}   & \multicolumn{1}{c}{}                  & \multicolumn{1}{l}{107} \\  
\multicolumn{1}{l}{ResNet101}   & \multicolumn{1}{c}{}                  & \multicolumn{1}{l}{209} \\  
\multicolumn{1}{l}{ResNet152}   & \multicolumn{1}{c}{}                  & \multicolumn{1}{l}{311} \\  
\multicolumn{1}{l}{MobileNet}   & \multicolumn{1}{c}{}                  & \multicolumn{1}{l}{55} \\ \hline
\multicolumn{1}{l}{MobileNetV2}   & \multicolumn{1}{c}{CIFAR-100}                  & \multicolumn{1}{l}{115} \\ \hline
\multicolumn{1}{l}{T5-small}   & \multicolumn{1}{c}{\multirow{5}{*}{ConLL2003}} & \multicolumn{1}{l}{50} \\  
\multicolumn{1}{l}{T5-large}  & \multicolumn{1}{c}{}                  & \multicolumn{1}{l}{194} \\ 
\multicolumn{1}{l}{T5-3B}  & \multicolumn{1}{c}{}                  & \multicolumn{1}{l}{194} \\
\multicolumn{1}{l}{Llama3.2-1B}  & \multicolumn{1}{c}{}                  & \multicolumn{1}{l}{146} \\  
\multicolumn{1}{l}{Phi-1.5}  & \multicolumn{1}{c}{}                  & \multicolumn{1}{l}{170} \\
\hline
\end{tabular}
\end{table}

\begin{figure*}[t!]
\centering
% EfficientNetB0
\includegraphics[width=1.0\textwidth]{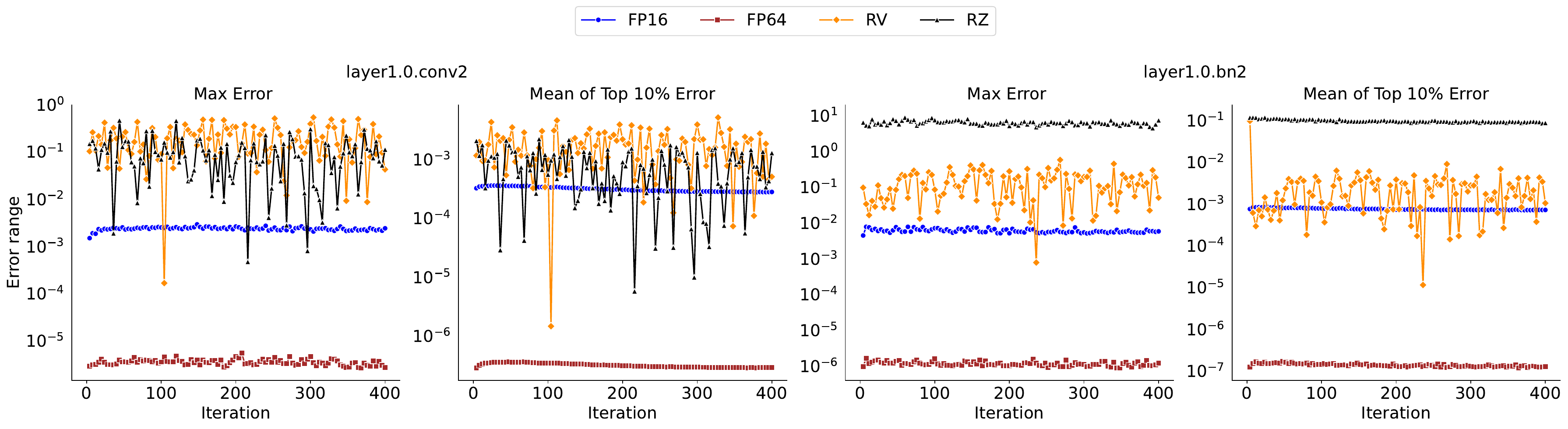}
\includegraphics[width=1.0\textwidth]{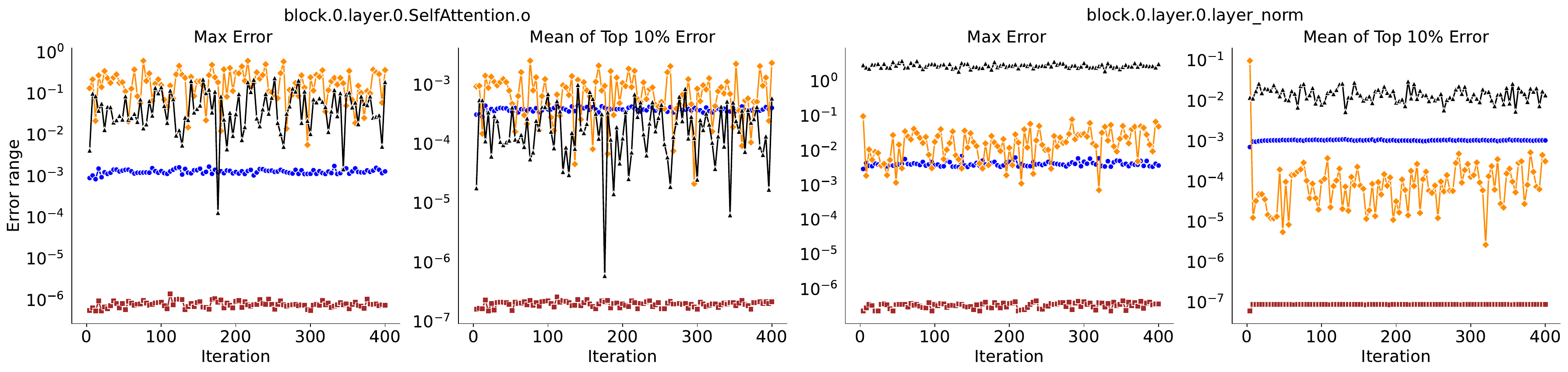}

\vspace{-5pt} % Adjust spacing with caption
\caption{Error ranges for ResNet18 (top) and T5-small (bottom) across different layers.}
\label{fig:profiler_error_ranges}
\end{figure*}

\subsection{Methodology}
\textbf{Workloads, Datasets, and Training Systems.} To demonstrate the effectiveness of PEAT, we validate it with eight popular CNN models, including Resnet-18/34/50/101/152~\cite{He2015_resnet}, MobileNet-v1 and v2~\cite{Andrew_MBv1, Mark_corr18_MBV2}, and EfficientNet-B0~\cite{tan2020_efficientnet}, and five well-known Transformer models, including T5-small, T5-large and T5-3B ~\cite{google_2020t5}, Phi-1.5~\cite{phi_gunasekar2023textbooksneed}, and Llama-3.2~\cite{touvron2023llama2openfoundation}. MobileNet-v2 is trained with the CIFAR-100 dataset ~\cite{alex_cifar10} while the other CNN models are trained with the CIFAR-10 dataset~\cite{alex_cifar10} for an image classification task. The Transformer models are trained and fine-tuned using the ConLL2003 dataset for an entity recognition task with English and German data. The experiments are conducted on NVIDIA Tesla V100S-PCIE-32GB~\cite{nvidia_v100} and AMD MI250~\cite{amd_mi250} GPUs. The detailed settings are summarized in Table~\ref{tab:training_models}.

\textbf{Error Models and Setups.} 
The code is implemented with PyTorch 1.13.1~\cite{paszke2019_pytorch}. 
We utilize two schemes to inject an error into a kernel during DNN training. 
One is to build an error kernel based on CUDA~\cite{cuda} for Nvidia GPUs and OpenCL~\cite{opencl} for AMD GPUs, enabling us to emulate a buggy code. 
The second scheme is to solely using PyTorch, which does not require any other libraries or packages. 
As mentioned in Section~\ref{sec:method}, we use four widely-known error models, including FP64, FP16, random value (RV), and random zero (RZ). 
%\rebuttal{Error models are used to collapsing a kernel input during forward, emulating a persistent calling of a faulty code during training.}
%\debug{For GEMM-like operations, we only focus on weight-activation layers, including CONV and FC layers in CNNs and FCs in Transformer models, where errors are injected into the weight tensors.} 
For RV and RZ, the number of error positions is empirically set to 1 for the Playback FI and 1, 3, or 5 for the frequency-based runtime FI. The error position is randomly selected. For the RV error model, an injected value at a pixel is randomly selected from the distribution of a target tensor.

\textbf{Fault Injection and Settings.} 
For a given model, we obtain the golden data by running its layers in FP32 with a fixed seed. 
Experiments with Playback FI are conducted with CONV, linear, and batch normalization layers in the CNN models and FC and layer normalization layers in the Transformer models, thanks to the low runtime overhead and the storage efficiency of the Playback FI. 
Additionally, frequency-based runtime FI experiments are performed on the first blocks in EfficientNet, Resnet-18, and T5-small. 
For RV and RZ, an error is injected within 25\%, 50\%, and 75\% of total epochs. 
For the Transformer models, we validate PEAT through both pretraining and fine-tuning tasks.

\begin{figure}[t!]
\centering
% EfficientNetB0
\includegraphics[width=\columnwidth]{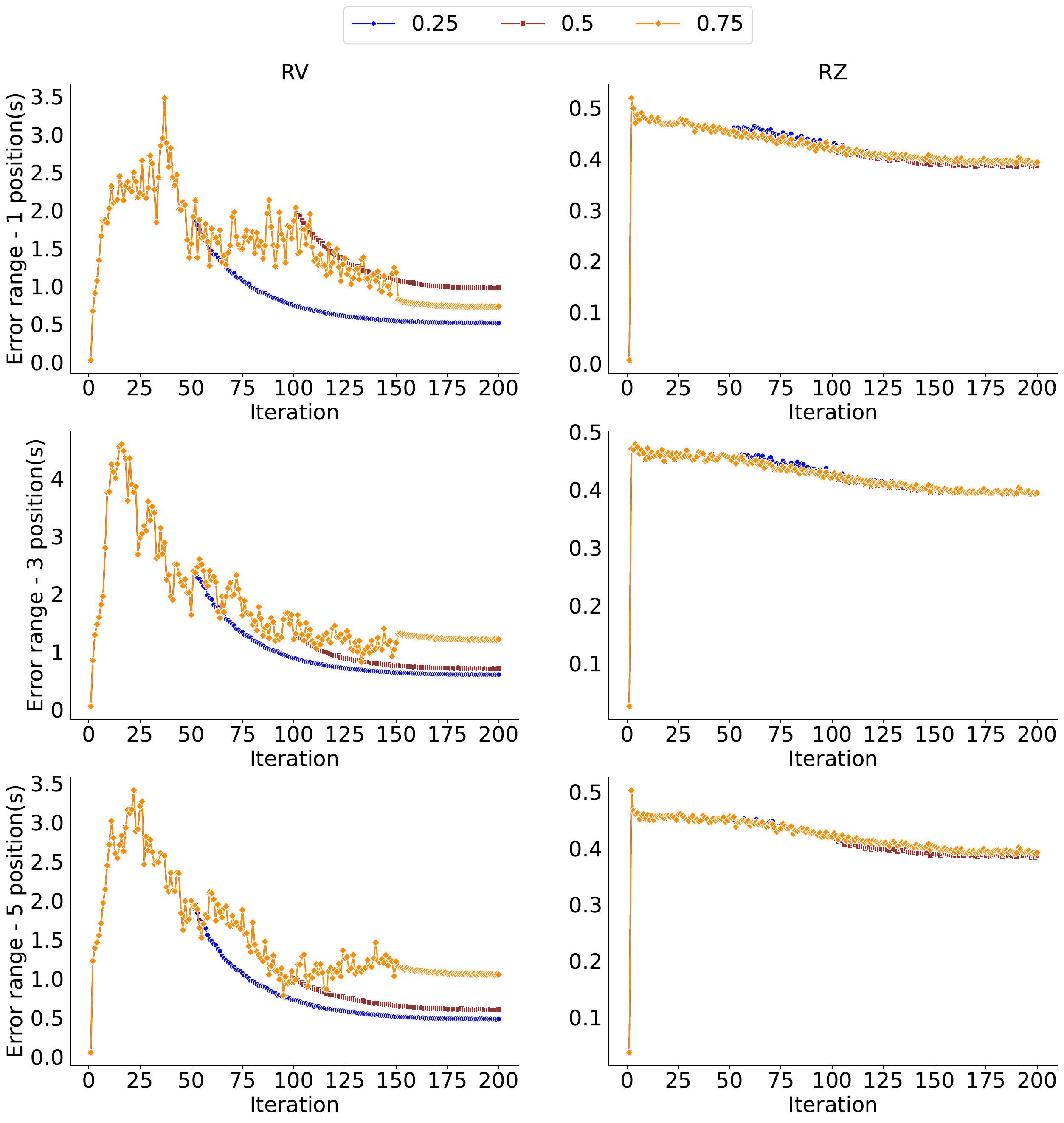}

\vspace{-5pt} % Adjust spacing with caption
\caption{Errors with Frequency-based runtime FI with different injected points: top (1), middle (3), and  bottom (5) for a CONV layer in EfficientNet.}
\label{fig:profiler_runtime_fi}
\end{figure}

%\subsection{Key Results and Discussions}
\subsection{Profiler Results}
%This section presents FI errors of fundamental operations in DNN models. 
%The experiment is conducted with CONV1 and BN1 in Resnet-18~\cite{He2015_resnet}, along with layer normalization and attention output FC layer in Block 1 of T5-small~\cite{google_2020t5}. 
%Results of other blocks are omitted due to their similar patterns.

\debug{
This section presents the error profile results.
For visualization, the results with Conv1 and BN1 in Resnet-18~\cite{He2015_resnet} and an attention output FC and LN layer in the first block of T5-small~\cite{google_2020t5} are presented as an example. 
}

\subsubsection{Playback FI} 
%\textbf{CNN models.} 
For CNN models, as shown in Fig.~\ref{fig:profiler_error_ranges}, FP16 and FP64 errors exhibit a similar trend, oscillating within a small change w.r.t. the training epoch. Meanwhile, RV and RZ errors show a significant variation between epochs, as illustrated in Fig.~\ref{fig:profiler_error_ranges} (top). 
As max and top-10\% mean values share a similar pattern, we can use a top-10\% score to represent an absolute error with a smoother variation. 
%Additionally, as discussed in Section~\ref{sec:method}, FP64 errors, considered as an error unit, are used to obtain a relative error. 
Meanwhile, for Transformer models, Fig.~\ref{fig:profiler_error_ranges} (bottom) shows the results for a Transformer-based model (T5). In general, we observe similar patterns, similar to those in CNN models. FC and CONV layers behave similarly, while LN and BN layers behave similarly. The main difference is that RV and RZ in LN layers exhibit smaller errors than in BN layers.
%\textbf{Transformer models.} 
%TRUONG -- Minh dang mo ta sai thi phai -> nen anh sua lai. Vi du FP16, FP64 khong thay tang khi epoch tang cho Transformers
%Similar to that of BN layers in a CNN model, FP16 and FP64 errors with LN layers in a Transformer-like model also tend to oscillate within a small change according to the training epoch, as shown in Fig.~\ref{fig:profiler_error_ranges} (bottom). 
%RV and RZ errors show a large variation among epochs.
%The main difference with CNN models is that FP16 and FP64 errors with FC layers show an increasing trend across epochs. 
%Fig.~\ref{fig:profiler_error_ranges} (bottom) shows the results for a Transformer-based model (T5). In general, we observe similar patterns as in CNN models. FC and CONV layers behave similarly, while LN and BN layers behave similarly. The main difference is that RV and RZ in LN layers exhibit smaller errors than in BN layers.

\subsubsection{Runtime FI} 
\label{subsubsec:Operation-wise FI Errors}
Fig.~\ref{fig:profiler_runtime_fi} reports RV (left) and RZ (right) errors, by injecting an error within 25\%, 50\% and 75\% of the total epochs for a CONV layer in EfficientNet.
In this figure, the top, middle, and bottom figures show the errors obtained with 1, 3, and 5 injected points. 
When an RV fault is injected over \quan{a specified number} of epochs, the obtained error generally remains large during that period and tends to decrease afterward.
Another substantial observation is that RV is more sensitive to the injection duration than RZ. RV errors show higher error ranges when the injection duration becomes longer (25\%, 50\% and 75\%).

\subsection{Analyzer Results}
%% Insight 1: Mean and Variance of error ranges
%%  Efficient-Net: CONV, T5: FCs

%% Insight 2: Roles of Batch and Layer Normalization 
%% Resnet-18, Phi

%% Insight 3: Repetitive patterns
%% 3a: Six residual blocks in Resnet-50
%% 3b: 3 encoder blocks in Llama-3.2
\subsubsection{Error Means and Variance}
\begin{figure}[t!]
\centering
\includegraphics[width=0.97\columnwidth]{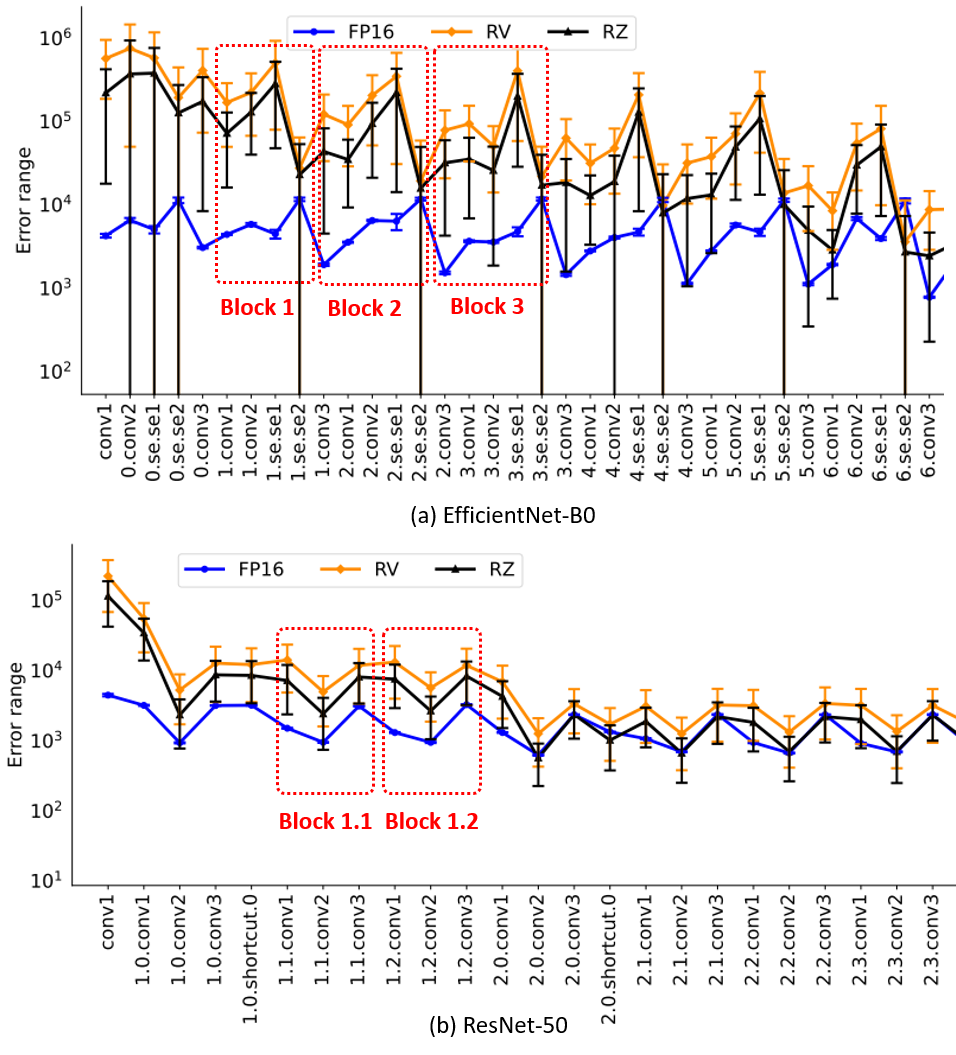}
\vspace{-10pt}
\caption{\nxt{Error Means and Variances with EfficientNet-B0 and ResNet-50.}}
\label{fig:analyzer_MBV2_ResNet50}
\end{figure}

\begin{figure}[t!]
\centering
\includegraphics[width=0.97\columnwidth]{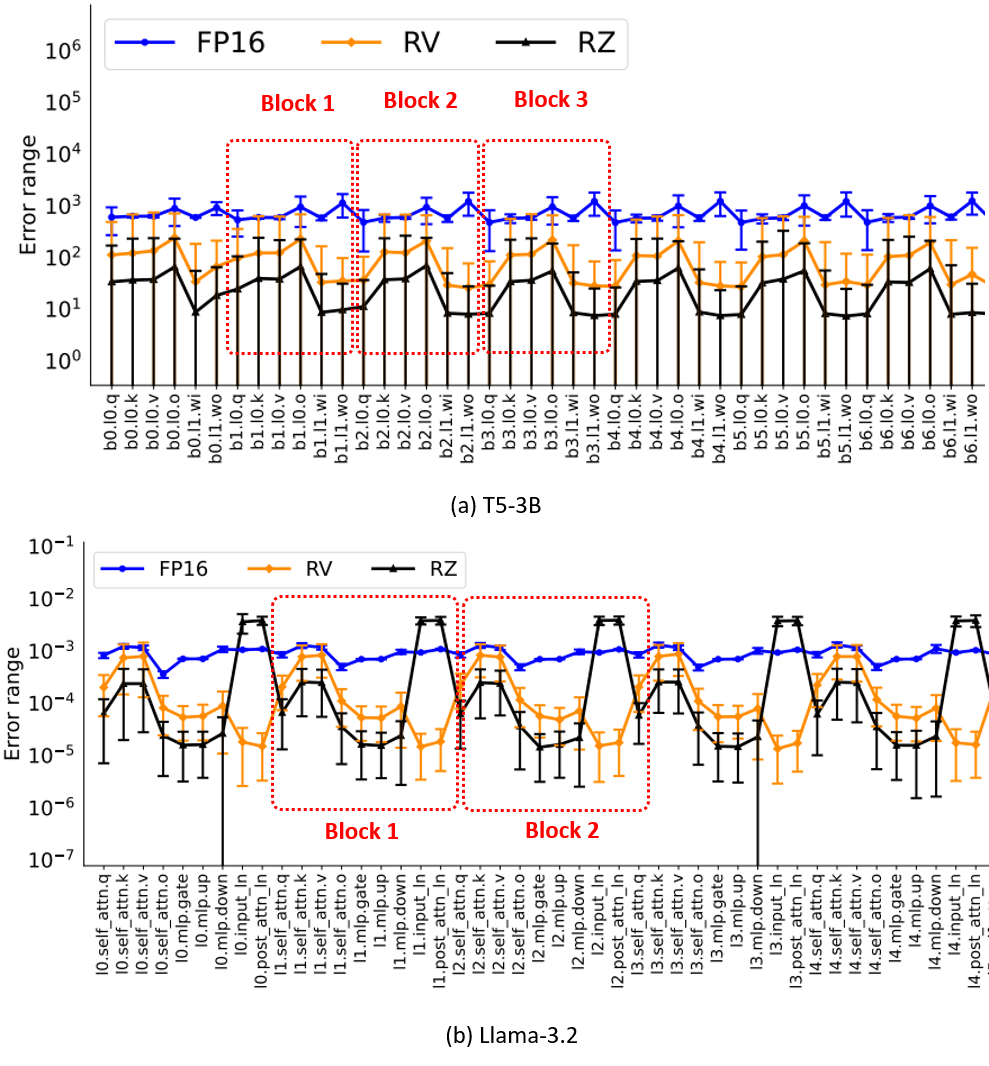}
\vspace{-10pt}
\caption{\nxt{Error Means and Variances with T5-3B and Llama-3.}}
\label{fig:analyzer_T5_Llama3}
\end{figure}

Fig.~\ref{fig:analyzer_MBV2_ResNet50} and Fig.~\ref{fig:analyzer_T5_Llama3} report the means and variances of relative errors with EfficientNet-B0, ResNet-50, T5-3B, and Llama-3. 
Here, the relative errors of FP16, RV, and RZ are obtained by dividing their corresponding absolute errors by their FP64 counterpart. 
The x-axis displays the layers, while the y-axis shows the error ranges represented by a pair of $\mu$ and $\sigma$. 
\final{As the errors are normalized, the ranges of errors in Fig.~\ref{fig:analyzer_MBV2_ResNet50} and Fig.~\ref{fig:analyzer_T5_Llama3} are much larger than those in Fig.~\ref{fig:profiler_error_ranges}. For example, as shown in in Fig.~\ref{fig:profiler_error_ranges}, the error ranges of FP16 (marked in blue) are in between $10^-3$ and $10^-4$, while the normalized error ranges of FP16 (marked in blue) in Fig.~\ref{fig:analyzer_MBV2_ResNet50} and Fig.~\ref{fig:analyzer_T5_Llama3} are in between $10^3$ and $10^4$, indicating that the errors of FP16 are larger than those of FP16 by $10^3-10^4$ times.}
\debug{
In analyses 1, 2, and 3, experiments are conducted by corrupting data during the forward pass only. 
Meanwhile, the analysis 4 compares error ranges with weight, input, and gradient error corruptions.
}

%% FIGURES
\textbf{\textit{Analysis 1 (Convergence Patterns with FP16 Errors)}:} 
% 2025.7.28 NXT edit
\final{Fig.~\ref{fig:analyzer_MBV2_ResNet50} and Fig.~\ref{fig:analyzer_T5_Llama3} show the means of variances errors for various layers in EfficientNet-B0, ResNet-50, T5-3B, and Llama-3. 
In CNN models, particularly \TODO{EfficientNet-B0} and Resnet-50, FP16 kernels (indicated by circle markers) show significantly smaller variances compared to the RV and RZ kernels, as illustrated in Fig.~\ref{fig:analyzer_MBV2_ResNet50}. 
Among these CNN layers, squeeze-and-excitation (SE) layers typically have higher variances than other layers (Fig.~\ref{fig:analyzer_MBV2_ResNet50} (a)). Meanwhile, the error ranges of FP16 layers in Resnet-50 are likely decreasing across image scales, as shown in Fig.~\ref{fig:analyzer_MBV2_ResNet50} (b).
In Transformer models such as T5-3B and Llama-3, FP16 kernels show higher error ranges but with much smaller variances compared to RV and RZ kernels for FC and LN layers, as illustrated in Fig.~\ref{fig:analyzer_T5_Llama3}.
}

%In CNN models, particularly \TODO{EfficientNet-B0} and Resnet-50, FP16 kernels (indicated by circle markers) show significant smaller variances compared to RV and RZ kernels, as illustrated in Fig.~\ref{fig:analyzer_MBV2_ResNet50}. 
%Among these CNN layers, squeeze-and-excitation (SE) layers typically have higher variances than other layers (Fig.~\ref{fig:analyzer_MBV2_ResNet50} (a)). 
%%There is a clear pattern in Fig.~\ref{fig:analyzer_MBV2_ResNet50} (a), where the errors of 
%%TRUONG -- CHECK THIS
%In Transformer models such as T5-3B and Llama-3, FP16 kernels show higher error ranges but with much smaller variances compared to RV and RZ kernels for FC and LN layers, as illustrated in Fig.~\ref{fig:analyzer_T5_Llama3}.

\textbf{\textit{Analysis 2 (Batch and Layer Normalization)}:} 
Interestingly, in CNN models, for BN layers, RZ (indicated by triangular markers) typically shows the highest error mean compared to the FP16 and RV errors, as demonstrated in Fig.~\ref{fig:analyzer_MBV2_ResNet50}. 
A similar pattern is also observed with LN layers in Transformer-based models, as illustrated in Fig.~\ref{fig:analyzer_T5_Llama3}. 
This suggests that normalization layers are more sensitive to RZ errors.

%%{{{

\textbf{\textit{Analysis 3 (Repetitive Patterns)}:} 
It is observed that repetitive blocks - that have identical input and output shapes - likely have similar error patterns. 
For example, blocks 1.1 and 1.2 in Resnet-50 show similar error patterns, as shown in Fig.~\ref{fig:analyzer_MBV2_ResNet50}. 
It is worth noting that these blocks consist of three layers, for example, conv1x1, conv3x3, and conv1x1, as explained in Section~\ref{subsec:DNN Model Scaling}. 
In Transformer models, encoder or decoder blocks also share similar error patterns, as shown in Fig.~\ref{fig:analyzer_T5_Llama3}.

\begin{figure}[t!]
\centering
\includegraphics[width=1.0\columnwidth]{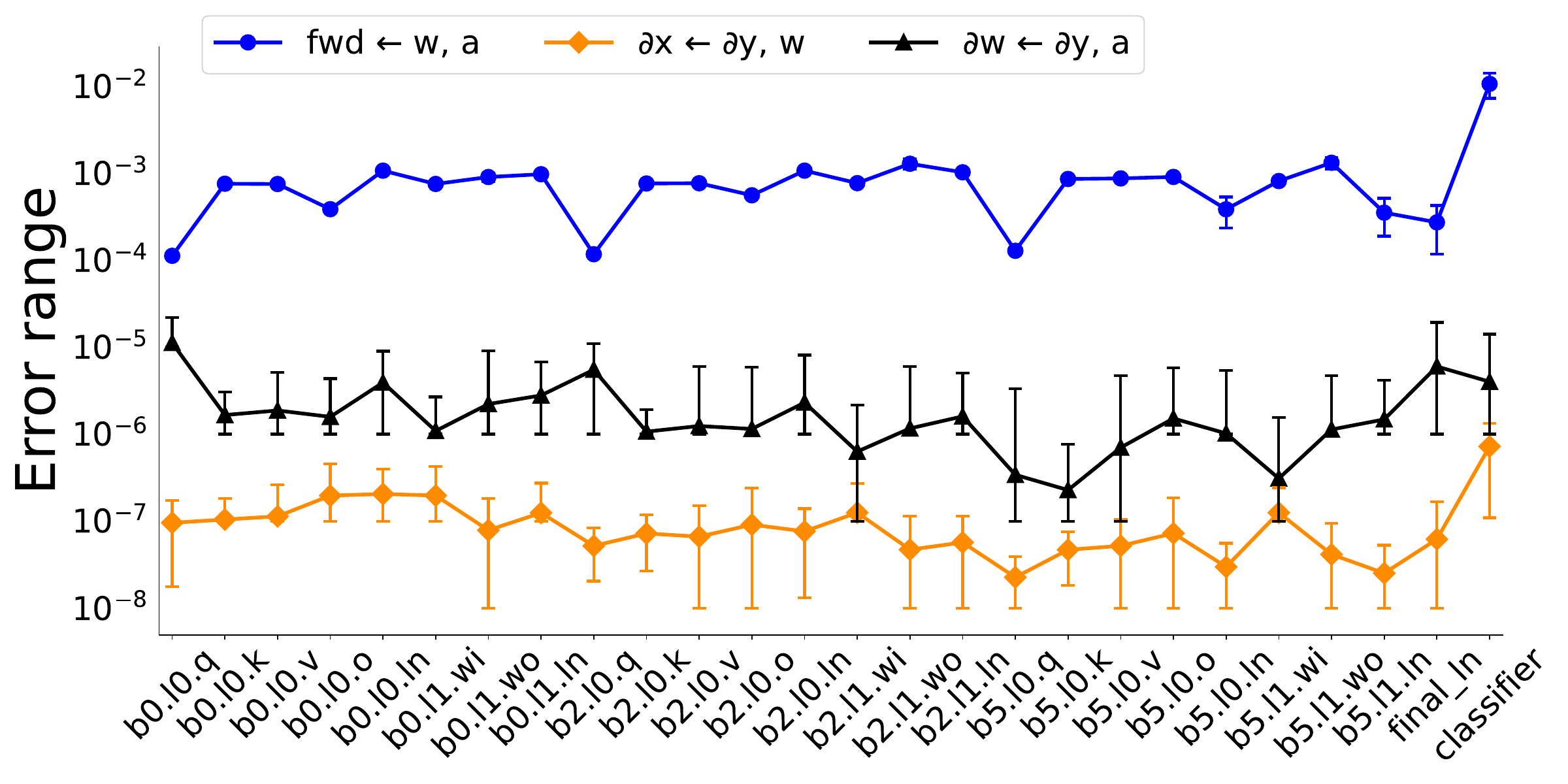}
\includegraphics[width=1.0\columnwidth]{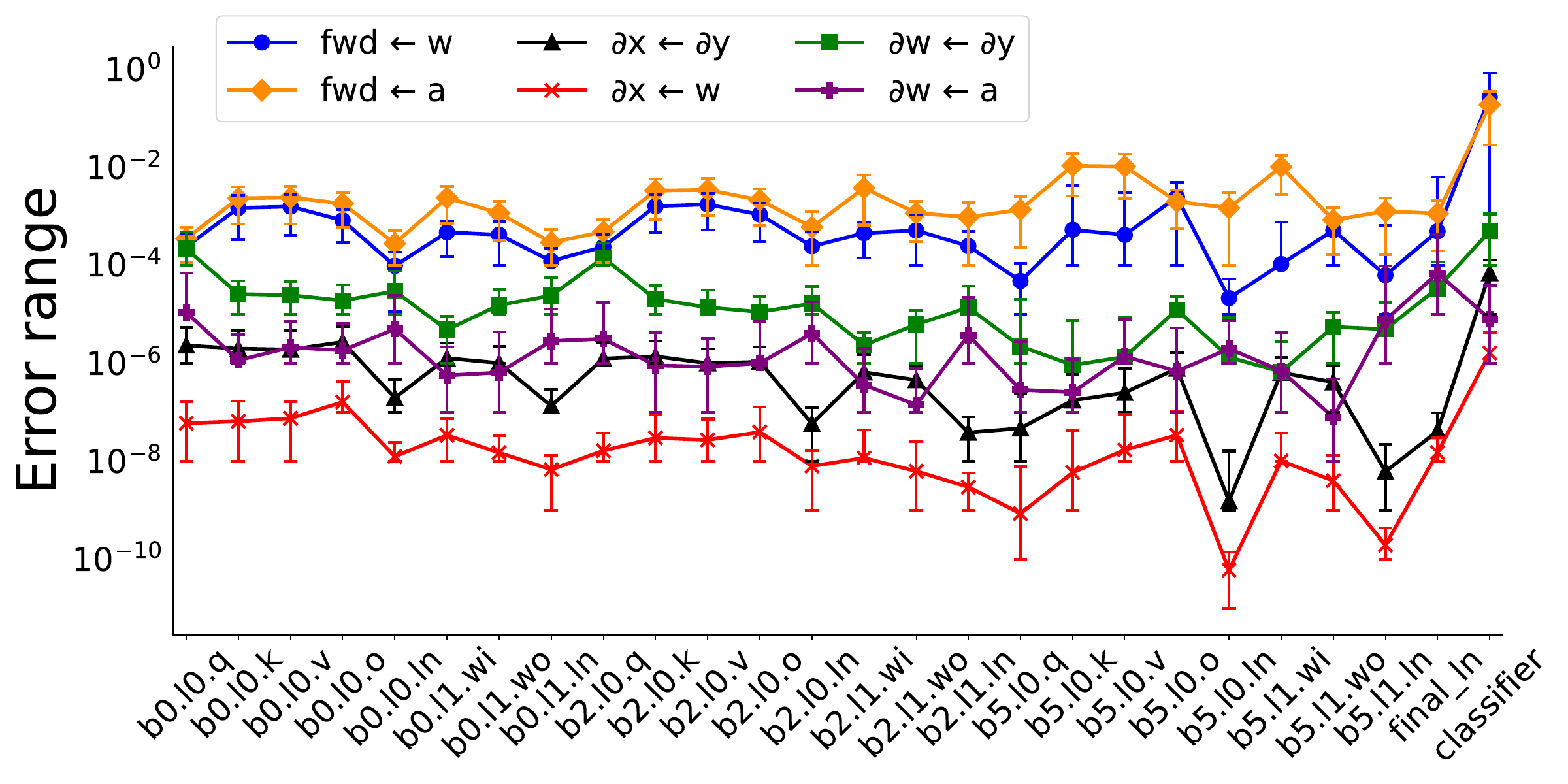}
\includegraphics[width=1.0\columnwidth]{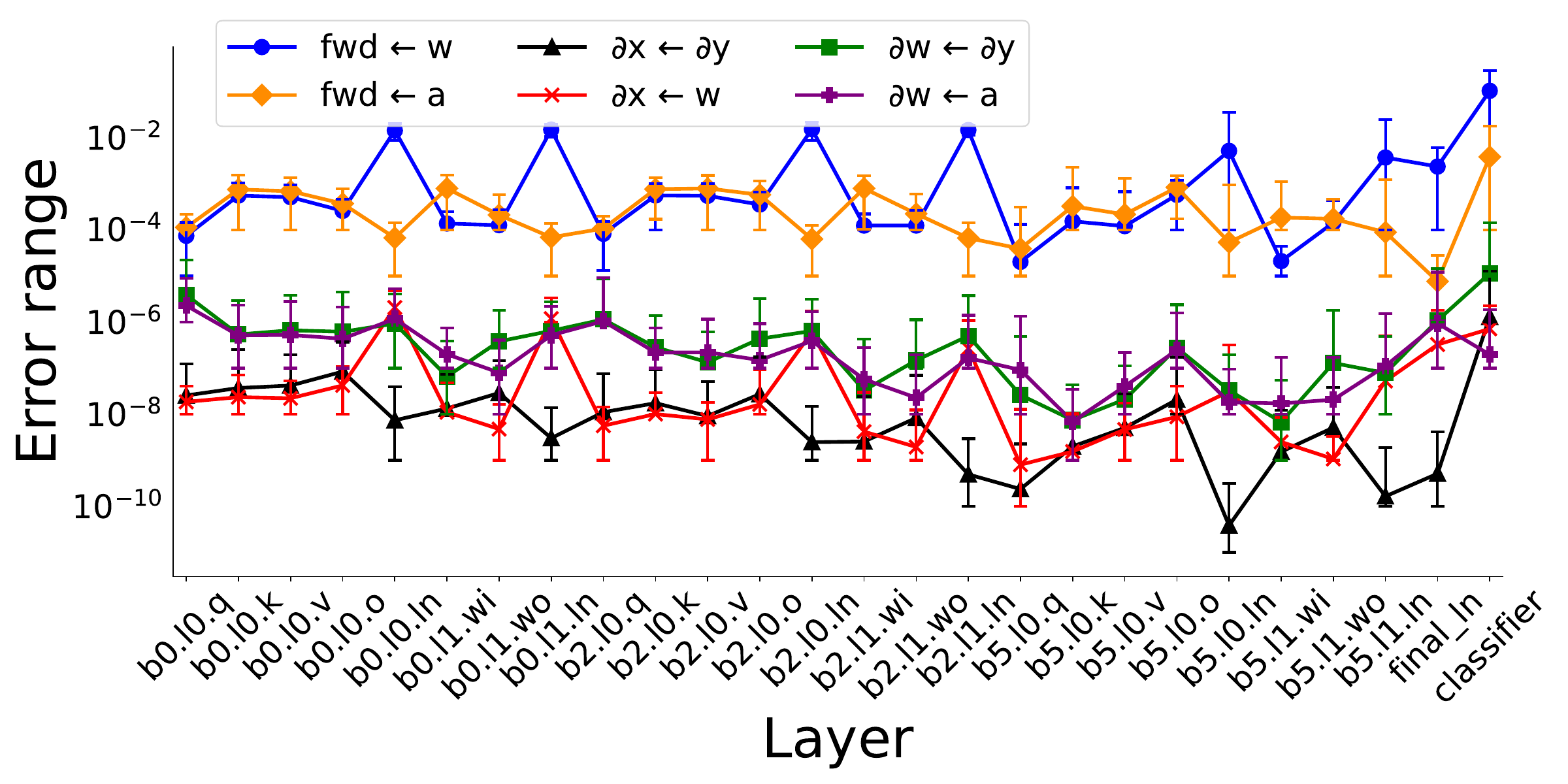}
\vspace{-10pt}
\caption{\nxt{Error Ranges on T5-Small (top: fp16; middle: rv; bottle: rz)}. \textit{fwd}$ \leftarrow \{w,a\}$ denotes injection location ($w$ or $a$) in the forward pass. 
$\delta x \leftarrow \delta y$ denotes computing gradient $\delta x$ and injection location at $\delta y$ in the backward.
}
\label{fig:analyzer_T5_small_forward_backward}
\end{figure}

\textbf{\textit{Analysis 4 (Corruption Locations)}:} 
\debug{
An unseen pattern is that corrupting a kernel in the forward pass introduces much larger errors than that in the backward pass.
For visualization, the results with the T5-small model are reported in Fig.~\ref{fig:analyzer_T5_small_forward_backward}. Patterns are consistently observed across blocks at different locations, for example, the first, middle, and later blocks.}

\begin{figure}[t!]
\centering
\includegraphics[width=1.0\columnwidth]{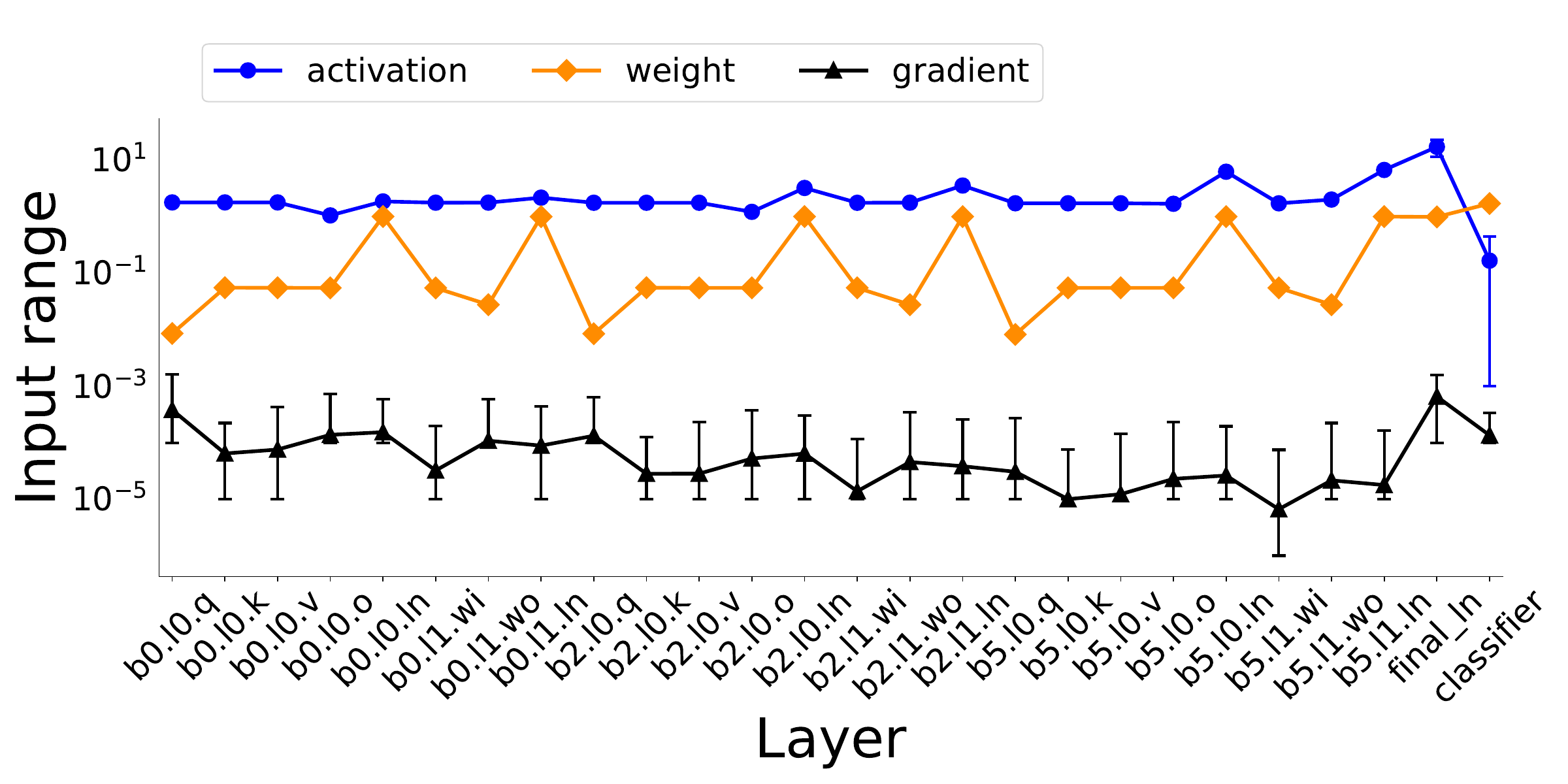}
\vspace{-10pt}
\caption{\nxt{Input Ranges of Activations, Weights, and Gradients in T5-Small.}}
\label{fig:analyzer_T5_small_forward_backward_input_range}
\end{figure}

\debug{Fig.~\ref{fig:analyzer_T5_small_forward_backward_input_range} shows the ranges of weights, activations, and gradients associated with the errors reported in Fig.~\ref{fig:analyzer_T5_small_forward_backward}. The range of gradients is significantly smaller than that of weights and activations during training. For particular dot product and matrix multiplication, the previous approaches such as ~\cite{Castaldo_siam08, Arar_siam23} have mathematically proved that an error bound depends on the ranges of inputs. However, our experimental results in Fig.~\ref{fig:analyzer_T5_small_forward_backward} show that practical errors are relatively small during DNN training, which cannot be seen by simply testing a kernel with random data.
}

%% }}}

%% }}}
\subsubsection{Correlation between Precision Errors and Other Factors}
\begin{figure}[t!]
\centering
% EfficientNetB0
\includegraphics[width=1.0\columnwidth]{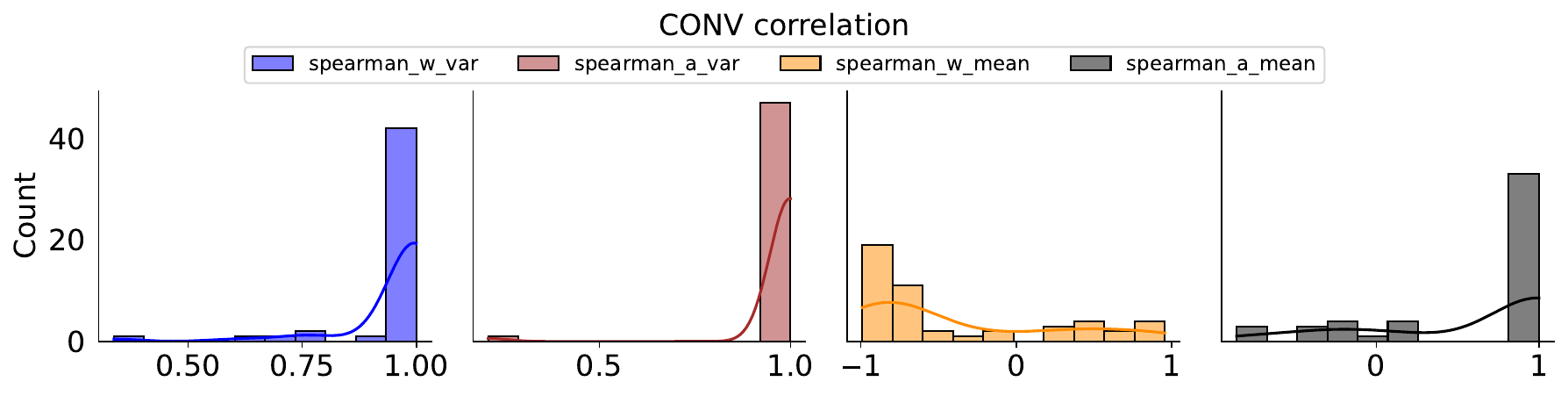}
\includegraphics[width=1.0\columnwidth]{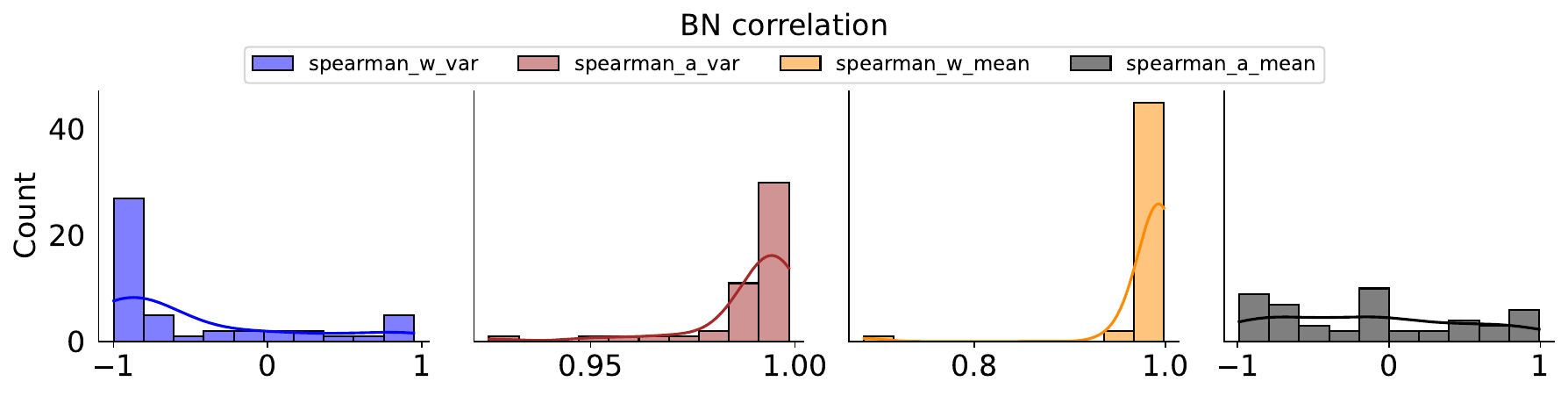}

\vspace{-5pt} % Adjust spacing with caption
\caption{Histogram of Spearman Scores for EfficientNet-B0 with CONV Layers (top) and BN Layers (bottom).}
\label{fig:profiler_spearman}
\end{figure}

%% Insight 4: Spearman > Signatures
%% 4a. Conv 
%% 4b. BN
%% 4c. FC
%% 4d. LN
\debug{
This section reveals the correlation between error patterns and other patterns, such as weight and activation means and variances. Fig.~\ref{fig:profiler_spearman} shows the Spearman score histogram of CNN and BN layers in EfficientNet-B0. It is observed that there is a strong correlation between a precision error and weight/activation variances, which is signified by high Spearman scores (ranging from 0.7 to 1.0). Meanwhile, for BN layers, the error and both the activation variances and the weight means show strong correlation with high Spearman scores. Accordingly, it suggests that errors can be monitored by visualizing other metrics such as activation variances or weight means.
}

%\textbf{Playback FI.} Fig.~\ref{fig:profiler_spearman} shows the Spearman score histogram of CNN and BN layers in EfficientNet-B0. 
%The scores quantify the relationship between an FP16 error and four criteria, including weight and activation means and variances. 
%As shown in Fig.~\ref{fig:profiler_spearman}(a), for CNN layers, there is a strong correlation between a precision error and weight/activation variances, which is signified by high Spearman scores (ranging from 0.7 to 1.0). 
%Meanwhile, Fig.~\ref{fig:profiler_spearman}(b) illustrates that in BN layers, the relationship between the error and both the activation variances and the weight means is significant. 
%These observations strongly suggest that it is possible to characterize precision errors by monitoring common metrics such as the means and variances of activations and weights.

\begin{figure}[t!]
\centering
% EfficientNetB0
\includegraphics[width=1.0\columnwidth]{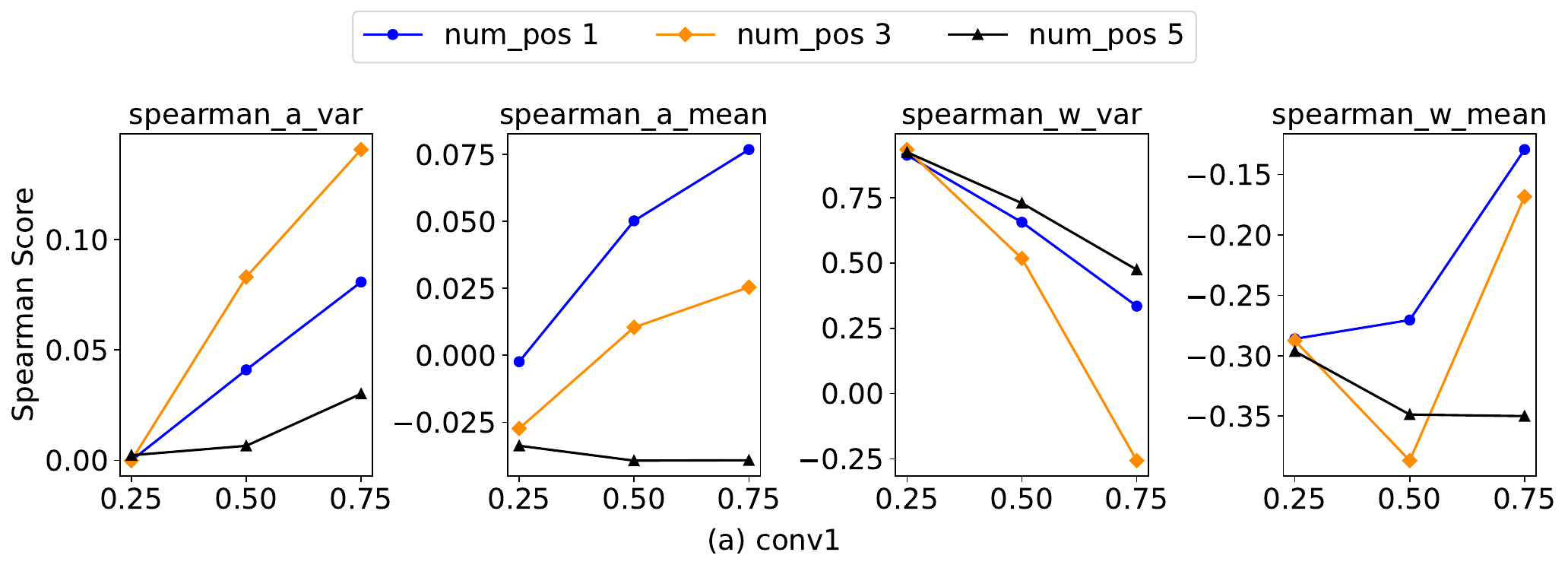}
\includegraphics[width=1.0\columnwidth]{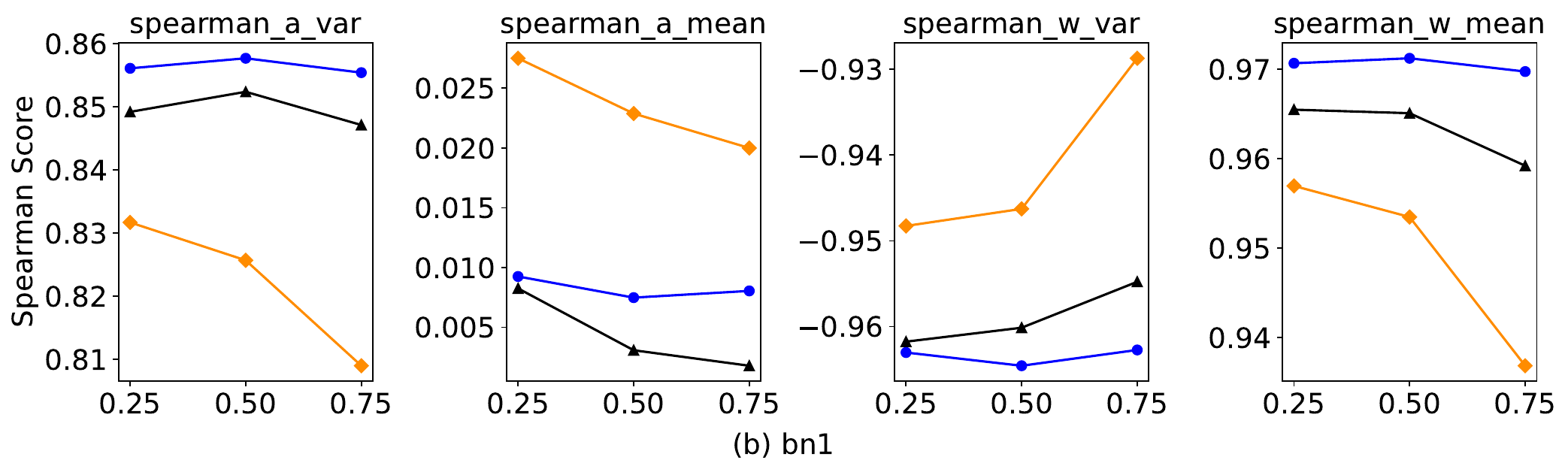}

\vspace{-5pt} % Adjust spacing with caption
\caption{\nxt{RV Spearman Scores on EfficientNet-B0.}}
%\caption{\TODO{Runtime Random Value Spearman Scores across ranges for layer CONV2 in the first block of EfficientNetB0.}}
\label{fig:profiler_spearman_runtime}
\end{figure}

%% Insight 5: Spearman > Signatures
%% Runtime FI
%\textbf{Runtime FI.} 
\debug{
Fig.~\ref{fig:profiler_spearman_runtime} shows the Spearman scores of Conv1 and BN1 layers in EfficientNet-B0 for runtime fault injection.
The experiment is conducted by injecting 1, 3, or 5 RV points within 25\%, 50\%, or 75\% of the total epochs. 
Like in the playback FI, it is observed that the relationship between the error and the activation variances and the weight means is also significant.
Additionally, Fig.~\ref{fig:profiler_spearman_runtime}(b) shows that the BN layer is relatively robust against RV errors because it shows a high Spearman score even in the worst-case scenario (e.g., injecting five positions over 75\% of the epochs). 
In contrast, CONV layers seem to be more sensitive to RV errors as the Spearman score largely drops for the 75\% case compared to the 25\% case.
}

%%{{{

%\debug{
%Fig.~\ref{fig:tiny_error_gemm_resnet34} reports an additional analysis of tiny errors when comparing a GEMM implementation with its golden version. 
%Synthetic data generally shows more significant errors than the training context data. 
%For example, the new GEMM kernel shows a maximum error with synthetic data, while that with training data shows maximum errors at iterations 0, 100, and 200, respectively.
%This experiment shows that different implementations during code optimization and porting can cause tiny errors.
%}

\subsubsection{Error Ranges with AMD MI250 GPU}
\begin{figure}[t!]
\centering
\includegraphics[width=1.0\columnwidth]{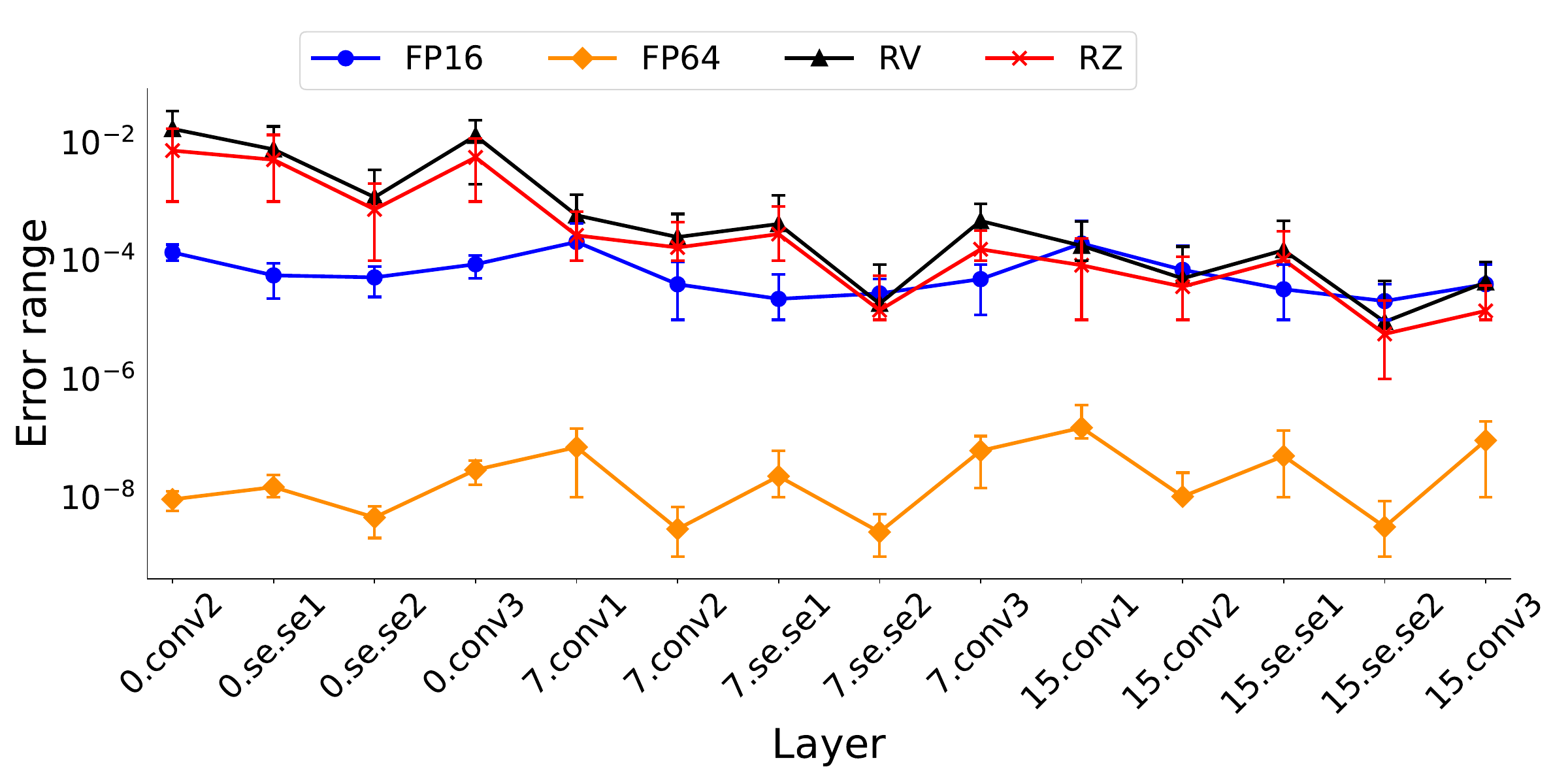}
\includegraphics[width=1.0\columnwidth]{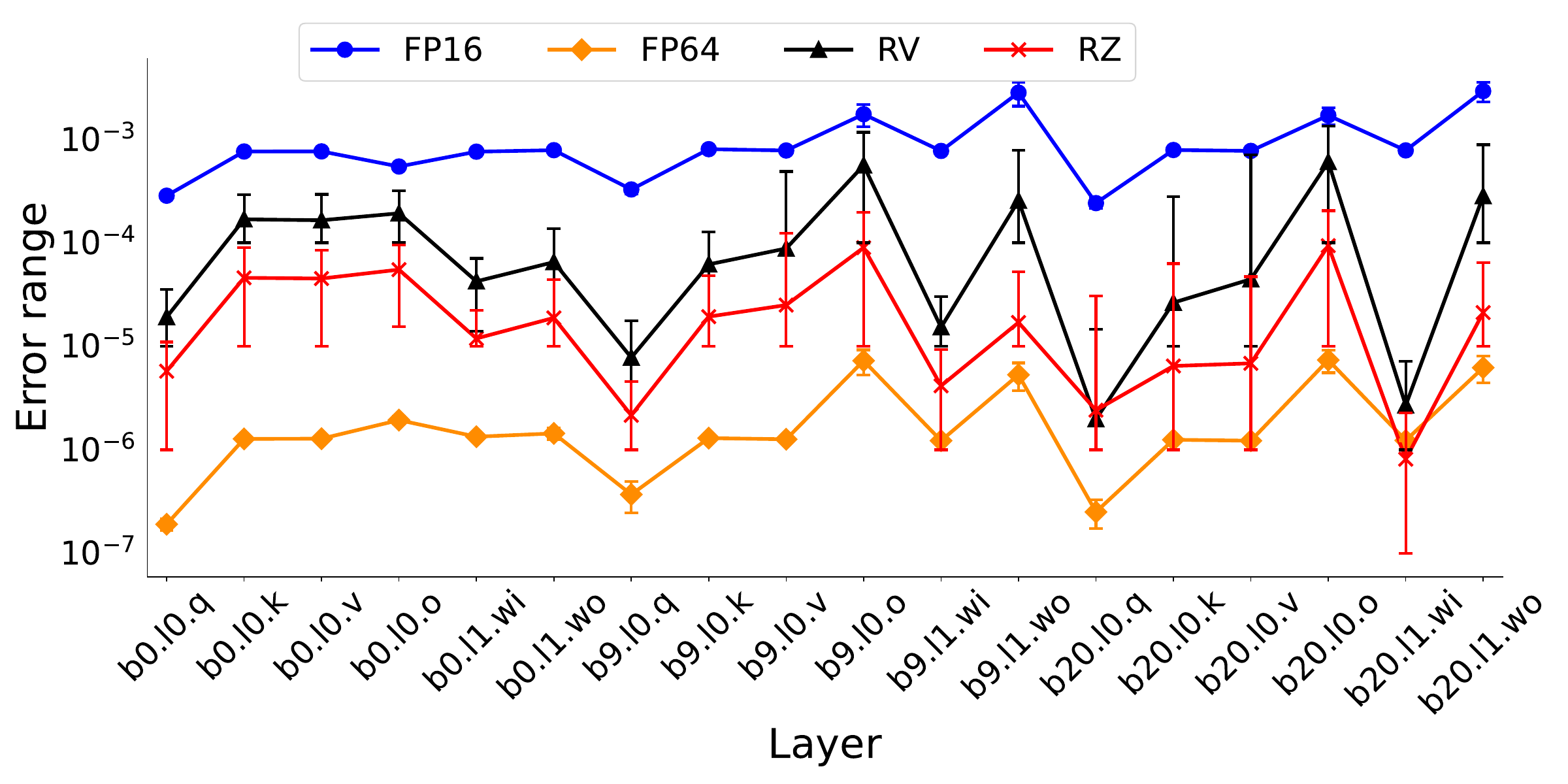}
% \vspace{-10pt}
\caption{\rebuttal{Error Ranges Measured with AMD MI250 GPU for EfficientNet-B0 (top) and T5-3B (bottom).}}
\label{fig:Insight1_Conv_amd}
\end{figure}

\debug{Fig.~\ref{fig:Insight1_Conv_amd} reports the error ranges measured on the AMD MI250 GPU ~\cite{amd_mi250} with EfficientNet-B0 and T5-3B. 
The results also highlight the convergence and repetitive patterns , similar to those with the Nvidia V100 GPU~\cite{nvidia_v100}.
\merge{For example, in EfficientNet-B0, each block contains three CONV layers and two SE-CONV layers, and the patterns are generally repeated every five layers. 
In MobileNet-v2, the error patterns are repeated every four layers, as illustrated in ~\ref{fig:Insight1_Conv_amd}.}} 
Interestingly, the means of RV and RZ errors appear to decrease with the model's depth.
Meanwhile, in common Transformer models~\cite{google_2020t5, touvron2023llama2openfoundation}, each encoder block consists of six FC layers, including Q, K, and V generation, a projection FC after self-attention, and two FCs in the feed-forward networks. 
In T5, the error patterns with FC layers tend to repeat every six layers.
Another key observation is that the errors across all three types - FP16, RV and RZ - are quite similar on both GPUs, enabling the effective application of the detection rules across these architectures.
This suggests that our algorithms do not rely on any hardware-specific assumptions, demonstrating wider applicability across different hardware and platform architectures.
%This enables the effective application of the detection rules across these architectures, demonstrating wider applicability across different hardware and platform architectures.}

%\begin{figure}[t!]
%\centering
%\includegraphics[width=1.0\columnwidth]{fig/Rebuttal/histogram_diff_ResNet341.pdf}
%% \vspace{-10pt}
%\caption{\rebuttal{Error Value Histograms with Synthetic Data and Training Context Data on Resnet-34.}}
%\label{fig:tiny_error_gemm_resnet34}
%\end{figure}

\subsection{\rebuttal{Tiny Errors, Training Loss, and Accuracy Drop}}

\merge{This section illustrates the impact of tiny errors on the resulted training loss and model accuracy.}

\subsubsection{\rebuttal{Tiny Errors}}
\begin{figure}[t!]
\centering
\includegraphics[width=1.0\columnwidth]{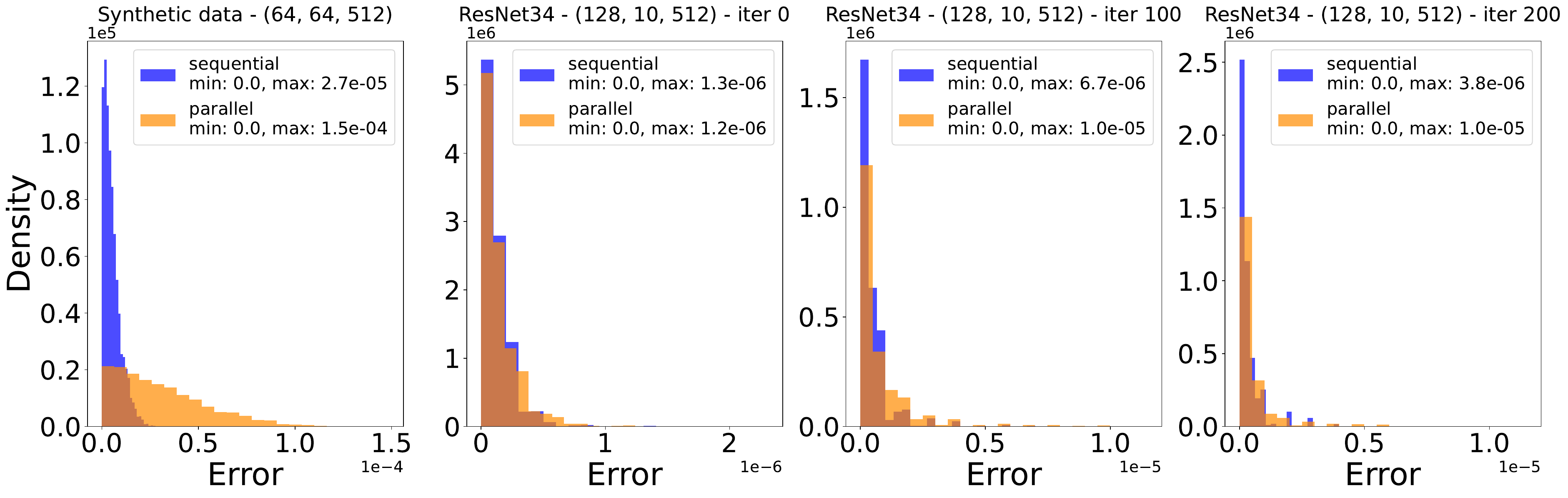}
% \vspace{-10pt}
\caption{\rebuttal{Tiny Error Analysis: Error value histograms with synthetic data and training context data on Resnet-34.}}
\label{fig:tiny_error_gemm_resnet34}
\end{figure}

\merge{Fig.~\ref{fig:tiny_error_gemm_resnet34} 
%and Fig.~\ref{fig:tiny_error_gemm_resnet50} 
reports an additional analysis of tiny errors when comparing a GEMM implementation with its golden version. 
Synthetic data generally shows larger errors than context data. For example, in Specifically, the new GEMM kernel shows a maximum error with synthetic data, while that with training data shows maximum errors at iterations 0, 100, and 200, respectively.}
This experiment shows that different implementations during code optimization and porting can cause tiny errors.

%\begin{figure}[t!]
%\centering
%\includegraphics[width=1.0\columnwidth]{fig/Rebuttal/histogram_diff_ResNet50.pdf}
%% \vspace{-10pt}
%\caption{\rebuttal{Tiny Error Analysis: : Error value histograms with synthetic data and training context data %on Resnet-50.}}
%\label{fig:tiny_error_gemm_resnet50}
%\end{figure}

\subsubsection{\rebuttal{Training Loss and Training Accuracy}}
\begin{figure}[t!]
\centering
\includegraphics[width=1.0\columnwidth]{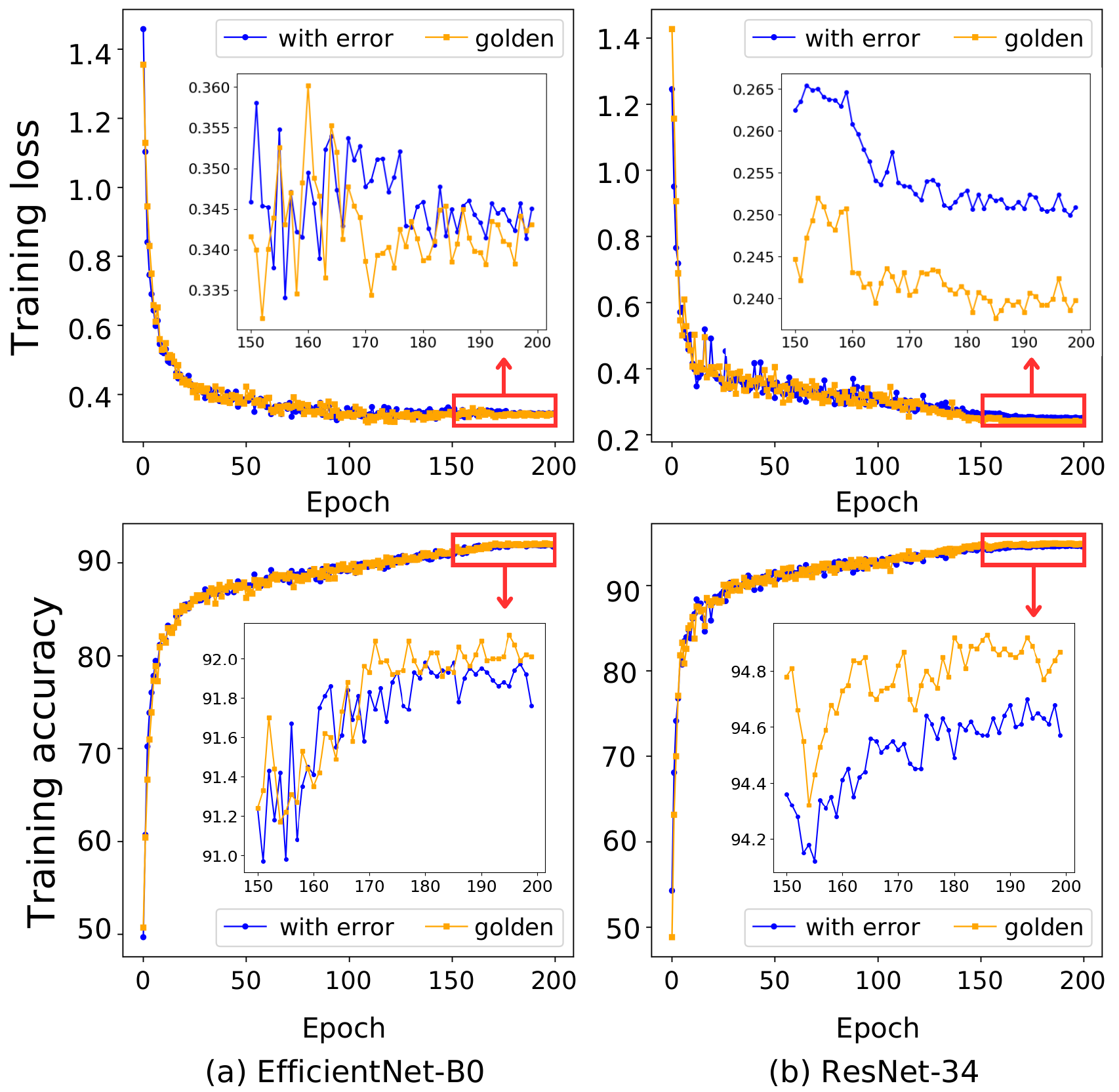}
% \vspace{-10pt}
\caption{Training accuracy drop with buggy GEMM kernels on CNNs}
\label{fig:cnn_loss_acc}
\end{figure}

\merge{Fig.~\ref{fig:cnn_loss_acc} presents additional results with training loss and training accuracy with EfficientNet-B0, Resnet-18, Resnet-34, and Resnet-50 when injecting an error in their GEMM kernels. 
%Fig.~\ref{fig:t5_loss_acc} presents additional results with training loss and training accuracy with EfficientNet-B0, Resnet-18, Resnet-34, and Resnet-50 when calling a new implementation of layer normalization in the T5 models. 
Although the errors observed during the kernel-level validation are very small, for example, the error range of $10^{-3}$, they often introduce noticeable degradation in the training loss and model accuracy.
}

%%%%%%%%%%%%%%%%%%%%%%%%%%%%%%%%%%%%%%%%%%%%%%%%%%%%%%%%%%%%%%%%%%%%%%%%%%%%%%%%%%%%%%%%%%%%%
%% DETECTION
%%%%%%%%%%%%%%%%%%%%%%%%%%%%%%%%%%%%%%%%%%%%%%%%%%%%%%%%%%%%%%%%%%%%%%%%%%%%%%%%%%%%%%%%%%%%%
%% TABLES
\begin{table*}[]
\caption{Error Range Statistics (CNNs)}
\label{tab:detector_cnn}
\vspace{-5pt}
\begin{center}
    \resizebox{0.85\textwidth}{!}{%
\begin{tabular}{ccllllllllllll}
\hline
\multirow{2}{*}{\textbf{Model}} & \textbf{Layer} & \multicolumn{4}{c}{\textbf{FP16}}                   & \multicolumn{4}{c}{\textbf{RV}}                        & \multicolumn{4}{c}{\textbf{RZ}}                      \\ \cline{2-14} 
                                & Operator       & xmin &xmax    & rmin& rmax    & xmin &xmax   & rmin &rmax      & xmin &xmax  & rmin& rmax     \\ \hline
\multirow{4}{*}{EfficientNet}   & CONV           & 2.89& 3.90 & 2.31& 6.01 & 3.09& 5.88  & 5.81& 11.70  & 2.66 &5.57 & 5.08& 11.51 \\ 
                              & FC             & 3.45 &3.46 & 4.50& 4.50 & 4.73& 4.73  & 9.20 &9.20   & 4.58 &4.58 & 8.43& 8.43  \\ 
                              & SE             & 3.56& 4.07 & 4.09 &6.68 & 3.00& 5.76  & 6.35& 11.55  & 2.83& 5.58 & 5.88 &11.16 \\ 
                              & BN             & 3.77& 3.82 & 4.72 &5.78 & 3.24 &5.90  & 6.84 &11.67  & 4.67& 6.47 & 6.57& 11.42 \\ \hline
\multirow{3}{*}{ResNet18}     & CONV           & 2.65&3.66 & 1.66& 4.54 & 2.22 &5.36  & 4.22 &10.41  & 2.02 &5.08 & 3.66& 9.76  \\ 
                              & FC             & 3.59& 3.59 & 4.81 &4.81 & 4.86 &4.86  & 9.48 &9.48   & 4.67& 4.67 & 8.94& 8.94  \\ 
                              & BN             & 3.72& 3.80 & 4.36& 4.98 & 2.24 &4.52  & 4.18& 8.85   & 4.96 &5.88 & 7.01 &9.60  \\ \hline
\multirow{3}{*}{ResNet34}     & CONV           & 2.59& 3.60 & 1.99 &4.45 & 2.17& 5.40  & 4.07 &10.41  & 1.71 &4.94 & 3.14& 9.66  \\ 
                              & FC             & 3.61 &3.61 & 5.24 &5.24 & 4.87 &4.87  & 9.50& 9.50   & 4.48 &4.48 & 8.73& 8.73  \\ 
                              & BN             & 3.72 &3.79 & 4.83 &5.55 & 3.38& 5.19  & 6.64& 10.36  & 4.89& 5.86 & 7.30 &10.14 \\ \hline
\multirow{3}{*}{ResNet50}     & CONV           & 2.54 &3.65 & 1.39& 4.52 & 1.90 &5.35  & 3.54 &10.35  & 1.63&5.06 & 2.91 &9.73  \\ 
                              & FC             & 3.90 &3.90 & 5.51& 5.51 & 4.45 &4.45  & 8.64 &8.64   & 4.13 &4.13 & 8.01 &8.01  \\ 
                              & BN             & 3.70& 3.79 & 3.89& 5.13 & 1.34& 4.41  & 2.34 &8.61   & 4.32 &5.87 & 5.47& 9.67  \\ \hline
\multirow{3}{*}{ResNet101}    & CONV           & 2.52& 3.50 & 1.88& 4.55 & 1.80 &5.24  & 3.32 &10.19  & 1.26 &4.66 & 2.38& 9.21  \\ 
                              & FC             & 3.90& 3.90 & 5.63& 5.63 & 4.21 &4.21  & 8.36& 8.36   & 3.83 &3.83 & 7.46& 7.46  \\ 
                              & BN             & 3.69 &3.80 & 4.72 &5.43 & 2.61 &5.45  & 5.05& 10.65  & 4.23 &5.87 & 6.73& 10.42 \\ \hline
\multirow{3}{*}{ResNet152}    & CONV           & 2.52 &3.47 & 1.71 &4.22 & 1.80& 5.00  & 3.34 &9.71   & 1.22 &4.50 & 2.35 &8.92  \\ 
                              & FC             & 3.91 &3.91 & 5.64 &5.64 & 4.23& 4.23  & 8.35 &8.35   & 3.87 &3.87 & 7.46& 7.46  \\ 
                              & BN             & 3.68& 3.80 & 4.68 &5.49 & 2.52& 5.75  & 4.91& 11.28  & 4.21& 5.87 & 6.62& 10.83 \\ \hline
\multirow{3}{*}{MBv1}         & CONV           & 2.89& 4.05 & 0.97& 5.04 & 2.38& 5.84  & 4.41& 11.43  & 2.14 &5.58 & 3.83& 10.90 \\ 
                              & FC             & 3.56& 3.56 & 4.61& 4.61 & 4.60& 4.60  & 8.88 &8.88   & 4.28& 4.28 & 8.25& 8.25  \\ 
                              & BN             & 3.72& 3.78 & 3.98 &5.58 & -0.07& 5.35 & -0.05& 10.86 & 4.69 &6.19 & 6.56 &10.55 \\ \hline
\multirow{3}{*}{MBv2}         & CONV           & 2.89& 3.91 & 1.58 &5.05 & 2.74& 5.79  & 5.16 &11.25  & 2.54 &5.61 & 4.63& 10.83 \\ 
                              & FC             & 3.98& 3.98 & 4.99 &4.99 & 3.79 &3.79  & 7.28& 7.28   & 3.32 &3.32 & 6.51 &6.51  \\ 
                              & BN             & 3.73& 3.81 & 3.96 &5.58 & 0.85& 5.36  & 1.63 &10.50  & 4.62 &6.47 & 6.20& 10.96 \\ \hline
\end{tabular}%
}
\end{center}
\end{table*}

\begin{table*}[]
\caption{Error Range Statistics (Transformers)}
\label{tab:detector_transformer}
\vspace{-5pt}
\begin{center}
    \resizebox{0.85\textwidth}{!}{%
\begin{tabular}{ccllllllllllll}
\hline
\multirow{2}{*}{\textbf{Model}} & \textbf{Layer} & \multicolumn{4}{c}{\textbf{FP16}} & \multicolumn{4}{c}{\textbf{RV}} & \multicolumn{4}{c}{\textbf{RZ}} \\ \cline{2-14} 
                             & Operator & xmin & xmax & rmin & rmax & xmin & xmax & rmin & rmax  & xmin  & xmax & rmin  & rmax  \\
\multirow{2}{*}{T5-small}    & FC       & 3.05 & 3.70 & 2.07 & 7.61 & 2.39 & 5.04 & 6.04 & 11.81 & 1.94  & 4.75 & 5.28  & 11.79 \\
                             & LN       & 4.04 & 4.09 & 8.05 & 8.61 & 3.03 & 4.29 & 7.37 & 11.09 & 4.96  & 5.24 & 10.76 & 13.16 \\ \hline
\multirow{2}{*}{T5-3B}       & FC       & 2.68 & 3.71 & 2.05 & 7.23 & 0.49 & 4.86 & 3.30 & 13.10 & -0.13 & 4.25 & 2.09  & 10.99 \\
                             & LN       & 4.02 & 4.10 & 7.86 & 8.74 & 2.00 & 3.23 & 4.75 & 9.72  & 4.39  & 4.99 & 9.54  & 13.13 \\\hline
\multirow{2}{*}{Llama3.2-1B} & FC       & 2.60 & 3.92 & 2.72 & 7.84 & 1.09 & 4.53 & 3.56 & 10.71 & 0.45  & 4.08 & 2.16  & 9.80  \\
                             & LN       & 4.19 & 4.26 & 7.22 & 7.99 & 2.34 & 0.27 & 7.08 & 8.36  & 4.75  & 4.81 & 10.57 & 11.89 \\\hline
\multirow{2}{*}{Phi-1.5}     & FC       & 2.69 & 3.92 & 3.72 & 7.87 & 1.46 & 4.49 & 3.77 & 10.44 & 0.89  & 4.04 & 2.80  & 9.59  \\
                             & LN       & 3.72 & 3.82 & 6.61 & 7.67 & 1.98 & 2.26 & 6.79 & 7.56  & 4.35  & 4.42 & 10.71 & 11.46 \\\hline
\end{tabular}%
}
\end{center}
\end{table*}

\subsection{Classification and Detection Results} \label{sec:Detection}
\debug{This section presents several key error patterns that enable the classification of error types and locations in kernel validation in DNN training. Table~\ref{tab:detector_transformer} report the maximum and minimum values of means and variances, denoted $[xmin, xmax]$ and $[rmin, rmax]$, respectively, during a training context of each model. We can monitor an error given by a new kernel code by extracting its error profile during the training context. The FP16, RV, and RZ errors are obtained by dividing them by the corresponding FP64 errors. Here are a few patterns: }

\debug{\textbf{Conv and FC Layers.} The variations of FP16 errors in Conv and FC layers are usually very small compared with those of RV and RZ. It is suggested that if an RV and RZ-like bug occurs in a kernel, testing it with a few epochs and several locations may reveal their large dynamic changes. Meanwhile, it is expected that an FP16-like correct kernel is likely to show a small variation across training epochs and various layers.}

\debug{ \textbf{BN and LN Layers.} Similarly, FP16 errors vary in a small range. Meanwhile, an RV-like bug in this layer is likely to show (1) relatively small changes in means (e.g., $(xmax-xmin) < th_1$) and (2) a high minimum range (e.g., $xmin > th_2$) across different locations. In Table~\ref{tab:detector_transformer}, we can empirically set $th_1=1.0$ and $th_2 = 10.0$ can detect RZ errors from RV and FP16-like errors.}

\debug{\textbf{Forward and Backward.} FP16-, RV-, and RZ-like errors in a kernel code show a consistent gap when corrupting a kernel during forward and backward recorded data. For example, an error by corrupting data in the backward pass is expected to be tiny. When it is noticeable, it can be further analyzed by corrupting data in the forward pass at various locations. In addition, the impact of this faulty kernel can be further explored by the runtime fault injection.}

%% file: sections/6_Related_works.tex
\section{Related Works} \label{sec:related work}
%\subsection{DNN Framework Development}
%% {{{
%Modern DNN training systems (???cite Pytorch, Pytorch-rocm, TensorFlow) and DNN inference systems (??? cite TVM, vLLM, MIGraphX) typically incorporate multiple GPUs and a system software stack, including a compiler and runtime. Users commonly work on domain-specific applications via popular Deep-Learning frameworks such as PyTorch~\cite{paszke2019pytorchimperativestylehighperformance} or TensorFlow~\cite{tensorflow2015-whitepaper}. 
%With the fast-paced development in AI workload, new DNN architectures often introduce new operations. 
%If the DNN architectures become sufficiently popular, the DNN framework developers will implement and optimize the new operations in various target hardware architectures. 
%Typically, what happens to be the common case is that the developers implement or optimize the operation code and compare the results for a set of operation inputs with that of a \textit{golden} version.
%The diversity in hardware architecture, the compiler optimization and kernel fusion together make the operation code validation a challenging task.
%Besides, each operation typically supports different modes, tackling different computation algorithms and execution precision.

\nxt{\textbf{DNN Framework Development.}} 
\merge{Modern DNN training systems (Pytorch~\cite{paszke2019_pytorch}, Pytorch-rocm, TensorFlow~\cite{tensorflow2015-whitepaper}) and DNN inference systems (TVM~\cite{Tianqi_tvm}, vLLM~\cite{kwon2023efficient}, MIGraphX~\cite{MIGraphX}) typically incorporate multiple GPUs and a system software stack, including a compiler and runtime. Users commonly work on domain-specific applications via popular DL frameworks such as PyTorch~\cite{paszke2019_pytorch} or TensorFlow~\cite{tensorflow2015-whitepaper}. 
With the fast-paced development of AI workloads, new DNN architectures~\cite{furiosa_tcp, SambaNova, Intel_Gaudi} often introduce new operations. The development of various target hardware architectures may demand an emerging trend of code development and optimization.
}

\nxt{\textbf{Kernel Code with Tiny Errors.}} 
\merge{A tiny error is highly related to precision errors due to floating-point computations, which is vastly studied in literature~\cite{Castaldo_siam08, blanchard_2019, fasi_2021_tensor_cores, Arar_siam23}. However, these approaches barely consider a practical DNN training context. For example, the proposed Playback FI reveals that the error mean of a convolution code tends to change in a small range during DNN training, which is underexplored in the conventional theoretical analysis~\cite{Castaldo_siam08, blanchard_2019, fasi_2021_tensor_cores, Arar_siam23}.
}

\nxt{\textbf{Error Characterization, Detection, and Mitigation.}} 
\merge{Soft errors have received a vast of attention recently in both DNN training systems~\cite{He_isca2023_HWFI, Guerrero_sc_23} and inference systems~\cite{Ma_asplos24_drDNA, Kamath_ISCA_Data_race, Bolchini_sc23, Singh_sc23}. A system fault during DNN training is widely known to be hard to detect. For example, according to ~\cite{He_isca2023_HWFI}, in 82.3\%$-$90.3\% of all cases across the workloads, the injected faults did not significantly affect the final training/test accuracy for the same training time as the fault-free runs. 
In fact, the majority of them (65.5\%$-$86.3\% of all cases) yielded slightly higher training/test accuracy compared to the fault-free cases. More importantly, it is costly to characterize a fault during DNN training, for example, $>$2.9 M ($>$490 K node hours) FI experiments~\cite{he_micro20_fidelity}.
PEAT, on the other hand, offers a low-complexity solution to investigate a new kernel under a DNN training context.
}

%% file: sections/7_Conclusion.tex
\section{Conclusion} \label{sec:conclusion}
We introduce PEAT, a kernel characterization framework, that is built on top of two simple yet effective FI techniques, Playback FI and frequency-based runtime FI. 
PEAT serves as a cost-effective tool to characterize and detect tiny errors in a kernel during the framework development process.
PEAT provides valuable insights for framework developers and simplifies device code validation by identifying the occurrences of potential stealthy errors in a specific DNN device code.
PEAT does not rely on hardware or platform-specific assumptions to function, allowing it to perform reasonably well across different development environments.
%Our framework can be seamlessly integrated with any other PyTorch-based framework, including Megatron-LM and DeepSpeed.
%PEAT can also be extended to multi-GPU and multi-node DNN training systems, considering heterogeneity communication in the training process.

% \cite{delebecque:hal-02090402}